\documentclass[11pt]{article}
\usepackage{jcappub,natbib}

\usepackage{cancel}

\title{Constraining primordial magnetic fields with
distortions of the black-body spectrum of the cosmic microwave
background: pre- and post-decoupling contributions}
\author[a]{Kerstin E. Kunze,}
\author[b,c]{Eiichiro Komatsu}
\affiliation[a]{Departamento de F\'\i sica Fundamental and IUFFyM, Universidad de Salamanca, Plaza de la Merced s/n, 37008 Salamanca, Spain}
\affiliation[b]{Max-Planck-Institut f\"ur Astrophysik, Karl-Schwarzschild Str. 1, 85741 Garching, Germany}
\affiliation[c]{Kavli Institute for the Physics and Mathematics of the
Universe, Todai Institutes for Advanced Study, the University of Tokyo,
Kashiwa, Japan 277-8583 (Kavli IPMU, WPI)} 
\emailAdd{kkunze@usal.es}
\emailAdd{komatsu@mpa-garching.mpg.de}
\abstract{%
Primordial magnetic fields that exist before the photon-baryon
decoupling epoch are damped on length scales below the photon diffusion
and free-streaming scales. The energy injected into the plasma by
dissipation of magnetosonic and Alfv\'en waves heats photons, creating a
$y$-type distortion of the black-body spectrum of the cosmic microwave
background. This $y$-type distortion is converted into a $\mu$-type
distortion when elastic Compton scattering is efficient. Therefore, we
can use observational limits on $y$- and $\mu$-type distortions to
constrain properties of magnetic fields in the early universe. Assuming
a Gaussian, random, and non-helical field, we calculate $\mu$ and $y$ as
a function of the present-day strength of the field, $B_0$, smoothed
over a certain Gaussian width, $k_c^{-1}$, as well as of the spectral
index of the power spectrum of fields, $n_B$, defined by $P_B(k)\propto
k^{n_B}$. For a nearly scale-invariant spectrum with $n_B=-2.9$ and a
Gaussian smoothing width of $k_c^{-1}=1~{\rm Mpc}$, the existing
COBE/FIRAS limit on $\mu$ yields $B_0<40$~nG, whereas the projected
PIXIE limit on $\mu$ would yield $B_0<0.8$~nG. For non-scale-invariant
spectra, constraints can be stronger. For example, for $B_0=1$~nG with
$k_c^{-1}=1~{\rm Mpc}$, the COBE/FIRAS limit on $\mu$ excludes a wide
range of spectral indices given by $n_B>-2.6$. After decoupling, energy
dissipation is due to ambipolar diffusion and decaying MHD turbulence,
creating a $y$-type distortion. The distortion is completely dominated
by decaying MHD turbulence, and is of order $y\approx 10^{-7}$ for a few nG
field smoothed over the damping scale at the decoupling epoch,
$k_{d,dec}\approx 290~(B_0/1~{\rm nG})^{-1}~{\rm Mpc}^{-1}$.  This 
contribution is as large as those of the known contributions
such as reionization at a redshift of $z\approx 10$ and virialized
objects at lower redshifts. The projected PIXIE limit on $y$ would
exclude $B_0>1.0$ and $0.6$~nG for $n_B=-2.9$ and $-2.3$, respectively,
and $B_0>0.6$~nG for $n_B\geq 2$. 
Finally, we find that the current limits on the optical depth to Thomson
scattering restrict the predicted $y$-type distortion to be $y\lesssim
10^{-8}$.
}
\begin{document}
\maketitle
\flushbottom

\section{Introduction}
The cosmic microwave background (CMB) is a nearly perfect
black body \cite{firas1,firas2,firas3}. The thermal spectrum
observed today is the result of its continuous evolution during the
history of the universe. In the early universe, $z\gtrsim 2\times 10^6$, a
black-body spectrum is established by bremsstrahlung \cite{is1,is2} and
double-Compton scattering 
\cite{ddz,bddz}. Subsequent injections of energy into the plasma in
$z\lesssim 2\times 10^6$ may
distort the black-body spectrum \cite{weymann,zeld1,sz}. 

When energy is injected into the plasma after $z\approx 2\times 10^6$,
heated electrons alter the photon spectrum via elastic Compton
scattering. Unlike 
bremsstrahlung and double-Compton scattering, elastic Compton scattering
does not alter the number of photons, and thus the equilibrium distribution
of photons after energy injection is a Bose-Einstein distribution with a
non-vanishing chemical potential, $\mu$. The photon occupation number
distribution is then given by $n(x)=(e^{x+\mu}-1)^{-1}$, where $x\equiv
h\nu/(k_BT)$ is a dimensionless frequency of photons.
If elastic
scattering is not efficient enough for the equilibrium distribution to be
reached, the distribution of photons takes on a
$y$-type distortion \cite{zeld1,sz} or an intermediate form between 
$\mu$- and $y$-type distortions \cite{cs,ks}, depending on the nature
and  epoch of energy injection.
 
Acoustic waves in the photon-baryon plasma before the photon-baryon
decoupling epoch are damped on length scales below the photon diffusion
scale (which is often called the ``Silk damping scale'')
\cite{silk,py,kaiser}. This damping injects 
energy into the plasma, creating a $y$-type distortion
\cite{sz2,sz3,daly}. This process can also be understood as a
superposition of black-body spectra with different temperatures
\cite{zeld2,ksc2}. Recent calculations show that only 1/3 of the
injected energy creates a $y$-type distortion, while the remaining 2/3 of the
energy raises the temperature of photons \cite{cks,ksc2,pz}. 

If magnetic fields exist before the decoupling epoch, they are similarly
damped \cite{jko1,sb}, creating $\mu$- and $y$-type distortions
\cite{jko2}. After the decoupling epoch, 
magnetic fields are damped by 
different mechanisms (ambipolar diffusion and decaying
magnetohydrodynamical (MHD) turbulence \cite{sesu}), again creating 
a $y$-type distortion. In this paper, we first reexamine the calculation of
$\mu$- and $y$-type distortions due to damping of magnetic fields in the
pre-decoupling era (section~\ref{sec:predecoupling}). We then present the
calculation of a $y$-type 
distortion as well as of the optical depth to Thomson scattering
due to damping of magnetic fields in the {\it
post}-decoupling era (section~\ref{sec:postdecoupling}). 
Finally, we conclude in section~\ref{sec:conclusions}. We use the
best-fit cosmological 
parameters from the {\sl WMAP} 9-year data only \cite{wmap9} for
the numerical calculations presented in this paper.

\section{Pre-decoupling era}
\label{sec:predecoupling}
\subsection{Damping of primordial magnetic fields}
While there is one acoustic (compressible) mode in a non-magnetized
plasma in the early universe, there are three modes in a magnetized
plasma. Two are compressible and are called ``fast'' and ``slow''
magnetosonic waves, while the other is incompressible and is called an
``Alfv\'en'' wave \cite{jko1,sb}. Magnetosonic waves perturb the
plasma density as 
well as the magnetic field, and are scalar modes in linear
perturbation theory. An Alfv\'en wave perturbs the  field but
does not perturb the density, and is a vector mode in linear
perturbation theory. 

The evolution of these modes is rather complex, depends on the ratios
of damping and driving terms, and is in general quite different from
damping of acoustic waves in a non-magnetized photon-baryon
plasma. Nevertheless, the analyses given in Refs.~\cite{jko1,sb} provide a
rather simple picture. For weak fields, the (time-dependent) oscillation
frequency of the 
fast mode is given by $\omega= \pm v_s(k/a)$, where $k$ is the comoving
wavenumber; $v_s=1/\sqrt{3(1+R)}$ is the sound speed
of the photon-baryon plasma; $R\equiv3\rho_b/(4\rho_\gamma)\propto a$;
$a$ is the Robertson-Walker scale
factor; and $\rho_b$ and $\rho_\gamma$ are 
energy densities of baryons and photons, respectively. (We take the
speed of light to be  
unity.) The fast mode damps in the
same way as the acoustic wave in a non-magnetized plasma, and thus its
damping wavenumber, $k_d$, is determined by  the inverse of the 
usual photon diffusion scale,
 $k_{\gamma}$; namely, the fast mode is damped by an
exponential factor of $\exp(-k^2/k_d^2)$ with $k_d=k_{\gamma}$, where
\cite{kaiser} 
\begin{equation}
 \frac1{k_{\gamma}^2}
\equiv \frac16\int
\frac{dt}{a^2(1+R)}l_\gamma\left(\frac{16}{15}+\frac{R^2}{1+R}\right).
\label{kga}
\end{equation}
Here, $l_\gamma\equiv 1/(\sigma_Tn_e)$ is the mean free path of
photons. In a deeply radiation-dominated era in which $R\ll 1$, we have 
$k_{\gamma}^{-2}\to (8/45)\int dt~ l_\gamma/a^2$.

The oscillation
frequencies of the slow and Alfv\'en modes are equal  to $\omega=\pm
V_A\cos\theta(k/a)$ (up to the leading
order in the field value), where 
$\theta$ is an
angle between the wave vector and the field direction, and
$V_A=\tilde{B}/\sqrt{1+R}$ is the 
Alfv\'en velocity with
\begin{equation}
\tilde{B}\equiv \frac{B}{\sqrt{16\pi\rho_\gamma/3}}\simeq 
3.8\times 10^{-4}\left(\frac{B_0}{1~{\rm nG}}\right).
\end{equation}
Here, $B_0a_0^2\equiv Ba^2$ is the present-day field value assuming 
magnetic flux freezing.
These modes also damp in a similar way as 
the fast mode, if their comoving wavelength is longer than
$l_\gamma/(aV_A\cos\theta)$. However, in the opposite limit in which their 
comoving wavelength is shorter than $l_\gamma/(aV_A\cos\theta)$,
these modes become {\it over-damped} and are nearly frozen. They
survive damping until their comoving wavelength becomes comparable to
$l_\gamma/a\ll l_\gamma/(aV_A)$. In this free-streaming regime, the slow
and Alfv\'en modes with the comoving wavelength shorter than
$v_A\cos\theta$ times the photon diffusion scale are damped, i.e.,
$k_d=k_{\gamma}/(v_A\cos\theta)$, where 
\begin{equation}
v_A\equiv \tilde{B}=V_A\sqrt{1+R},
\end{equation}
 is the Alfv\'en velocity calculated during the radiation-dominated era (in
which $R\ll 1$). 

The longest comoving damping wavelength (i.e., smallest $k_d$) is given
 by that at the decoupling epoch, $z=z_{\rm dec}=1088$.
Modeling the recombination history following Ref.~\cite{HuSu} and using
 the exact expression of the photon diffusion scale given in
 eq.~(\ref{kga}), we find
\begin{eqnarray}
k_{d,dec}
=\frac{286.91}{\cos\theta}\left(\frac{B_0}{1~\rm nG}\right)^{-1}\;\; {\rm Mpc}^{-1},
\label{kdast}
\end{eqnarray}
for the  best-fit $\Lambda$CDM model from the {\sl WMAP} 9-year data
only (see Appendix \ref{sec:appendix} for details).
The wavenumbers of the slow and Alfv\'en modes contributing to the
observed $\mu$- and $y$-type distortions are {\it larger} than this
value.\footnote{Note that only the photon diffusion scale (the
Silk damping scale) is important for distortions of the CMB spectrum. 
In the evolution of a primordial magnetic field,  neutrino decoupling marks another important epoch for damping of a field.
Similarly to photon free-streaming, neutrino free-streaming damps
magnetosonic and Alfv\'en waves, injecting energy into the plasma. 
The corresponding damping scale is given in Ref.~\cite{sb}.
 However, neutrino decoupling takes place at $z\sim
 10^{10}$, at which time the elastic as well as double-Compton scattering
 is very effective, and a black-body spectrum is restored. Therefore, an
 energy injection due to dissipation of a field by neutrino
 free-streaming does not cause any observable distortion of the CMB
  spectrum.}

\subsection{Energy injection rate}
We calculate the energy injection rate, $\dot{Q}$, as a rate at which
the field energy density changes due to damping of magnetosonic
and Alfv\'en waves: $\dot{Q}=-a^{-4}d(\rho_Ba^4)/dt$. 

Let us define the comoving magnetic energy density, $\rho_{B,0}$, as 
\begin{eqnarray}
\rho_{B,0}(\vec{x})a_0^4\equiv\rho_B(\vec{x},t)a^4,
\end{eqnarray}
where the subscript 0 indicates the present time. In our units,
$\rho_{B,0}(\vec{x})=\frac{1}{2}\vec{B}^2_0(\vec{x})$, where
$B_0a_0^2\equiv Ba^2$. 

We assume that a field is a non-helical and Gaussian random
field with the two-point correlation function in Fourier space given by
\begin{eqnarray}
\langle B_i^*(\vec{k})B_j(\vec{q})\rangle=(2\pi)^3\delta({\vec{k}-\vec{q}})P_B(k)\left(\delta_{ij}-\frac{k_ik_j}{k^2}\right),
\end{eqnarray}
where the power spectrum, $P_B(k)$, is assumed to be a power law,
$P_B(k)=A_Bk^{n_B}$, with the amplitude, $A_B$, and the spectral index, $n_B$.

As the energy density of fields, $\rho_{B,0}$, is proportional to
$B_0^2$, the ensemble average of $\rho_{B,0}$ is given by the power
spectrum as
\begin{eqnarray}
\langle\rho_{B,0}\rangle=\int\frac{d^3k}{(2\pi)^3}P_{B,0}(k)e^{-2\left(\frac{k}{k_c}\right)^2}, 
\end{eqnarray}
where $k_c$ is a certain Gaussian smoothing scale. Using $\rho_{B,0}$
instead of $A_{B,0}$, we write the present-day power spectrum as
\begin{eqnarray}
P_{B,0}(k)=\frac{4\pi^2}{k_c^3}\frac{2^{(n_B+3)/2}}{\Gamma\left(\frac{n_B+3}{2}\right)}\left(\frac{k}{k_c}\right)^{n_B}\langle\rho_{B,0}\rangle.
\end{eqnarray}
The ``scale-invariant'' case corresponds to $n_B\rightarrow-3$, in which the
contribution to the energy density per logarithmic wavenumber,
$d\ln\langle\rho_{B,0}\rangle/d\ln k\propto
k^3P_{B,0}(k)e^{-2(k/k_c)^2}$, is independent of wavenumbers (up to
smoothing). Spectral indices of fields generated during
inflation are usually negative \cite{tw}, whereas they are positive for fields generated by causal processes such as the electroweak phase transition \cite{ew1,ew2,ew3, rev1,rev2,rev3,rev4}. Moreover, in the latter case $n_B$ has to be an even integer and  $n_B\geq 2$ \cite{dc}.

For the calculation of the energy injection rate,
$\dot{Q}=-a^{-4}d(\rho_Ba^4)/dt$, we take the smoothing
scale to be the damping scale, $k_d$, and obtain
\begin{eqnarray}
\nonumber
a^4\langle
\rho_{B}\rangle(z)&=&a^4\int\frac{d^3k}{(2\pi)^3}P_B(k)e^{-2\left(\frac{k}{k_d(z)}\right)^2}\\
&=&
 a^4_0\int\frac{d^3k}{(2\pi)^3}P_{B,0}(k)e^{-2\left(\frac{k}{k_d(z)}\right)^2}.
\label{eq:energy}
\end{eqnarray}
The energy injection rate is thus found to be 
\begin{eqnarray}
\frac{\frac{dQ}{dz}}{\rho_{\gamma}}=\frac{n_B+3}{2}
\left(\frac{\rho_{B,0}}{\rho_{\gamma,0}}\right)k_c^{-(n_B+3)}k_d(z)^{n_B+5}\frac{d}{dz}k_d^{-2}(z), 
\label{eq:rate0}
\end{eqnarray}
where we have used
$\Gamma[(n_B+5)/2]=[(n_B+3)/2]\Gamma[(n_B+3)/2]$, and 
$\rho_{\gamma}$ and $\rho_{\gamma,0}$ are the photon energy densities at a
given time and the present time, respectively.
We have removed $\langle\rangle$ from $\rho_{B,0}$ for
a simpler notation. 

Let us write the damping scale as $k_d=\alpha k_{\gamma}$, where
$\alpha=1$ for the fast magnetosonic mode and 
$\alpha=v_A^{-1}\simeq 2.6\times 10^3~(1\,{\rm nG}/B_0)$ for the slow
and Alfv\'en modes.\footnote{In principle, we should perform
the integration over angles in eq.~(\ref{eq:energy}) taking into account
the fact that the damping
scale for the slow 
and Alfv\'en modes depends on angles as $k_d=k_{\gamma}/(v_A\cos\theta)$. We shall ignore this subtlety and take
$\cos\theta=1$. This simplification provides a lower limit to the energy
injection rate.} 
For $n_B\ge -3$, the energy injection rate by
dissipation of the slow and  Alfv\'en modes is greater than that of the
fast mode. As we consider only $n_B\ge -3$ in this paper, we shall
use $\alpha=v_A^{-1}$ for the rest of this paper. Finally, we shall
assume energy equipartition of the fast, slow, and Alfv\'en modes, which will
introduce an additional factor of 2/3 to eq. (\ref{eq:rate0}).
Note that this is a non-trivial assumption: if, for some
reason, the slow and Alfv\'en modes are highly suppressed relative to
the fast mode, our analysis does not apply.

Deep inside the radiation-dominated era, $z\gg z_{eq}=3265$ \cite{wmap9},
the photon diffusion scale is well approximated by
$k_{\gamma}^{-2}=A_{\gamma}^2z^{-3}$ with $A_{\gamma}^2=5.9807\times
10^{10}\;{\rm Mpc}^2$  for the best-fit parameters from the {\sl WMAP}
9-year data only  (see Appendix~\ref{sec:appendix} for details).
The energy injection rate thus becomes (after multiplying
eq.~(\ref{eq:rate0}) by $2/3$ for equipartition)
\begin{eqnarray}
\frac{\frac{dQ}{dz}}{\rho_{\gamma}}=-(n_B+3)
\left(\frac{\rho_{B,0}}{\rho_{\gamma,0}}\right)\left(\frac{\alpha A_{\gamma}^{-\frac{1}{2}}}{k_c}\right)^{n_B+3}\left(1+z\right)^{\frac{3n_B+7}{2}}.
\label{eq:rate1}
\end{eqnarray}
For $n_B<-7/3$, the energy injection rate decreases with
redshifts; thus, the largest contribution comes from the lowest
redshift under consideration. For $n_B>-7/3$, the largest contribution
comes from the highest redshift under consideration, i.e., $z\approx
2\times 10^6$, above which the distribution becomes a Planck
distribution.

\subsection{CMB distortions}
There are three main processes determining the final CMB spectrum:
elastic Compton scattering, double Compton scattering, and
bremsstrahlung. Ref. \cite{ddz} finds that double Compton scattering
dominates over bremsstrahlung in the relevant epoch.

At $z\gtrsim 2\times 10^6$, double Compton scattering is efficient and
photons have a Planck distribution. At $z\lesssim 2\times 10^6$, double
Compton scattering is no longer efficient enough to erase signatures of
energy injection. However, as long as elastic Compton scattering is
efficient, the distribution approaches a Bose-Einstein distribution with
a non-vanishing $\mu$. This is the so-called ``$\mu$-era,'' which occurs
between redshifts of $5\times 10^4\lesssim z\lesssim 2\times 10^6$
\cite{ksc}. (Strictly 
speaking, the distortion takes on an intermediate form between $\mu$-
and $y$-type distortions in $1.5\times 10^4\lesssim z\lesssim 2\times
10^5$ \cite{cs,ks}, but we shall ignore this subtlety.)
During the subsequent ``$y$-era,'' elastic Compton scattering is no
longer efficient and the photon spectrum cannot relax to a
Bose-Einstein distribution; thus, the distortion remains a
$y$-type. 

The time evolution of $\mu$ is determined by \cite{hs}
\begin{eqnarray}
\frac{d\mu}{dt}=-\frac{\mu}{t_{DC}(z)}+\frac{1.4}{3}\frac{\frac{dQ}{dt}}{\rho_{\gamma}},
\end{eqnarray}
where a factor of $1/3$ in the second term accounts for a recent finding
that only $1/3$ of 
the energy injection contributes to spectral distortions
\cite{cks,ksc2,pz}.\footnote{The same factor of $1/3$ applies to
energy injection by dissipation of magnetosonic and Alfv\'en waves. The
magnetic field is coupled to electrons. Upon dissipation it heats
electrons, raising the electron temperature, $T_e$. Photons and electrons
are tightly coupled so that $T_{\gamma}=T_{e}$, i.e., any change in the
electron temperature is  
transmitted to the photon temperature. Therefore, energy injection due
to dissipation of the magnetosonic and Alfv\'en waves is  similar to that of the
acoustic waves in a non-magnetized photon-baryon fluid: injection of the
magnetic field energy leads to a mixture of
black body spectra for which it has been shown that 1/3 of the injected
energy leads to spectral distortions and 2/3 to raise the average
temperature.}
Here, $t_{DC}$ is the time scale for double Compton scattering,
\begin{eqnarray}
t_{DC}=2.06\times 10^{33}\left(1-\frac{Y_P}{2}\right)^{-1}\left(\Omega_bh^2\right)^{-1}z^{-\frac{9}{2}}\;\;{\rm s},
\end{eqnarray}
where $Y_P=0.24$ is the primordial helium mass abundance. The solution
of this equation during the radiation era is 
\begin{eqnarray}
\mu= \frac{1.4}{3}\int_{z_1}^{z_2}dz~ \frac{\frac{dQ}{dz}}{\rho_{\gamma}}e^{-\left(\frac{z}{z_{DC}}\right)^{\frac{5}{2}}},
\label{mu}
\end{eqnarray}
where the integration is done for the $\mu$-era, i.e., 
$z_1=2\times 10^6$ and $z_2=5\times 10^4$, and
\begin{eqnarray}
z_{DC}\equiv 1.97\times 10^6\left[1-\frac{1}{2}\left(\frac{Y_P}{0.24}\right)\right]^{-\frac{5}{2}}\left(\frac{\Omega_b h^2}{0.0224}\right)^{-\frac{2}{5}}.
\end{eqnarray}

Since the $\mu$-era is well within the radiation-dominated era, the
energy injection rate is given by eq.~(\ref{eq:rate1}). Thus,
\begin{eqnarray}
\mu&=&-\frac{1.4}{3}(n_B+3)
\left(\frac{\rho_{B,0}}{\rho_{\gamma,0}}\right)\left[\frac{1.08\times 10^{-2}\left(\frac{B_0}{\rm nG}\right)^{-1}}{k_c/{\rm Mpc}^{-1}}\right]^{n_B+3}
\nonumber\\
&&\times\int_{z_1}^{z_2}dz~ (1+z)^{\frac{3n_B+7}{2}}e^{-\left(\frac{z}{z_{DC}}\right)^{\frac{5}{2}}},
\end{eqnarray}
with $\frac{\rho_{B,0}}{\rho_{\gamma,0}}=9.545\times
10^{-8}\left(\frac{B_0}{\rm nG}\right)^2$ for $T_{\rm cmb}=2.725$~K \cite{pdg}.

For $z<z_2$, elastic Compton scattering is no longer effective, and thus
the distortion remains a $y$-type. The Compton $y$-parameter is given by
\cite{cks}
\begin{eqnarray}
y=\frac{1}{12}\int_{z_2}^{z_{\rm
 dec}}dz~\frac{\frac{dQ}{dz}}{\rho_{\gamma}}, 
\label{y}
\end{eqnarray}
where $z_{\rm  dec}=1088$ is the decoupling epoch \cite{wmap9}.
In this era the universe is matter-dominated, and we use the 
full expression for the photon diffusion scale with the recombination
history given by Ref.~\cite{HuSu}.
We shall discuss the post-decoupling contributions
separately in section~\ref{sec:postdecoupling}.

\subsection{Results}
There are three free parameters: the present-day field value, $B_0$,
smoothed over a certain Gaussian width of $k_c^{-1}$, and 
the spectral index, $n_B$. Among these, $n_B$ and $k_c$ should be
determined by a mechanism by which fields are
generated.\footnote{Alternatively, one
may simply take $k_c$ to be a convenient normalization scale at which
limits on $B_0$ are reported.} 

For fields generated during inflation, $k_c$ can take on any values,
while $n_B$ required to produce strong enough fields on cosmological
scales is usually negative.  The correlation length of fields generated by
causal processes such as a phase transition should be determined by the
horizon size of the relevant phase transition (QCD or electroweak),
while $n_B$ is positive and causality requires $n_B$ to be even integers
with $n_B\ge 2$ \cite{dc}. 

For numerical calculations we use $Y_P=0.24$ and $\Omega_bh^2=0.02264$
\cite{pdg,wmap9}. 

\subsubsection{$\mu$-type distortion}
\begin{figure}
\centerline{\epsfxsize=2.9in\epsfbox{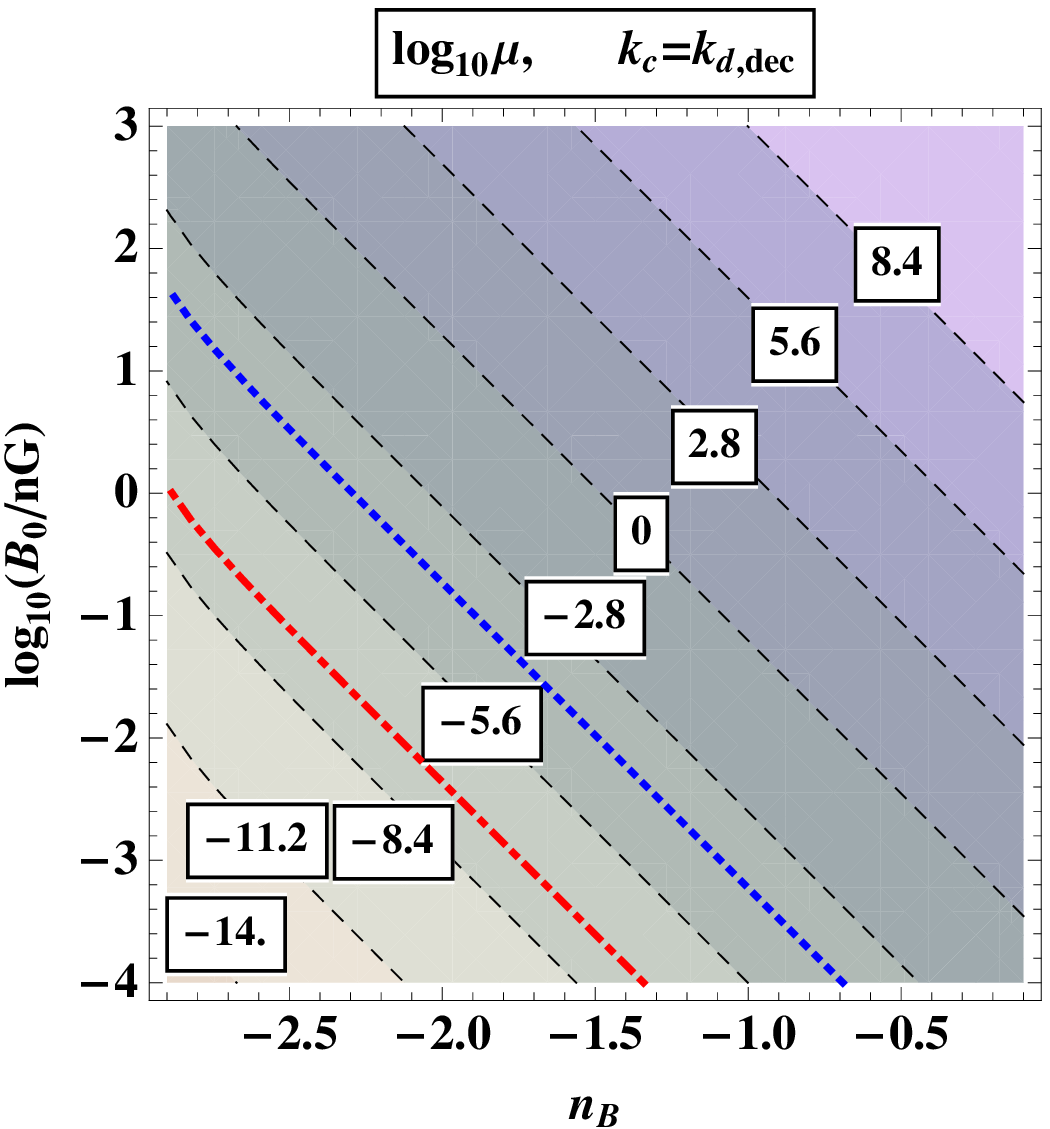}
\hspace{0.9cm}
\epsfxsize=2.9in\epsfbox{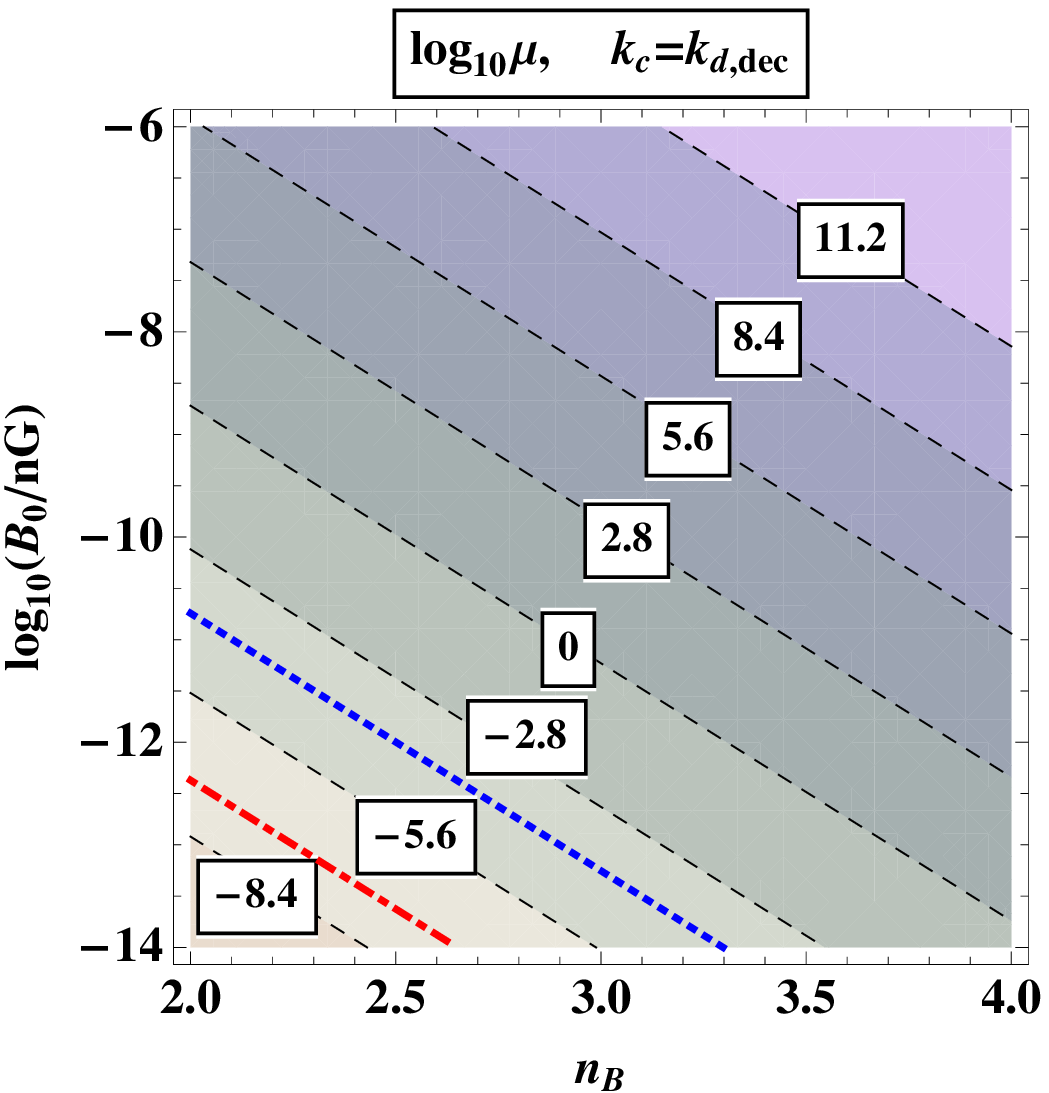}}
\caption{Contour plots of $\log_{10}\mu$ in the ($B_0$, $n_B)$
 plane for $n_B$ expected from inflation (left panel) and phase
 transitions (right panel).
 The smoothing scale is set to be the maximal 
 value of the damping wavelength at decoupling (eq.~(\ref{kdast}) with
 $\cos\theta=1$).   
The blue (dotted) lines show the COBE/FIRAS limit, $\mu=9.0\times
 10^{-5}$ \cite{firas3}, and the red (dot-dashed) lines show the 
projected PIXIE limit, $\mu=5.0\times 10^{-8}$ \cite{pixie}.}
\label{fig1}
\end{figure}

First, we set $k_c^{-1}$ to be the maximal damping wavelength at
decoupling (eq.~(\ref{kdast}) with $\cos\theta=1$). 
Figure~\ref{fig1} ({\it left}) shows the predicted values of $\mu$ as a
function of $B_0$ and $n_B$. The left and right panels show $n_B$
expected from inflation and phase transitions, respectively.

We find that the COBE/FIRAS limit, $|\mu|<9\times 10^{-5}$ \cite{firas3}
(blue dotted lines), yields $B_0<40 $~nG for a nearly
scale-invariant spectrum, $n_B=-2.9$. 
The projected PIXIE limit, $|\mu|<5\times 10^{-8}$ \cite{pixie} (red
dot-dashed lines), would yield an order of magnitude stronger
constraint, $B_0<0.9$~nG, for $n_B=-2.9$. 
For non-scale-invariant spectra, $n_B>-2.9$, the constraints are much
stronger. For example, the COBE/FIRAS limit and the projected PIXIE
limit yield $B_0<0.4$ and $10^{-2}$~nG, respectively, for
$n_B=-2.2$. For $n_B\ge 2$ expected from phase transitions, the largest
$B_0$ allowed by the COBE/FIRAS limit is of the order of 
$10^{-11}$~nG, which can be improved to $3\times10^{-13}$~nG using PIXIE.

\begin{figure}
\centerline{\epsfxsize=2.9in\epsfbox{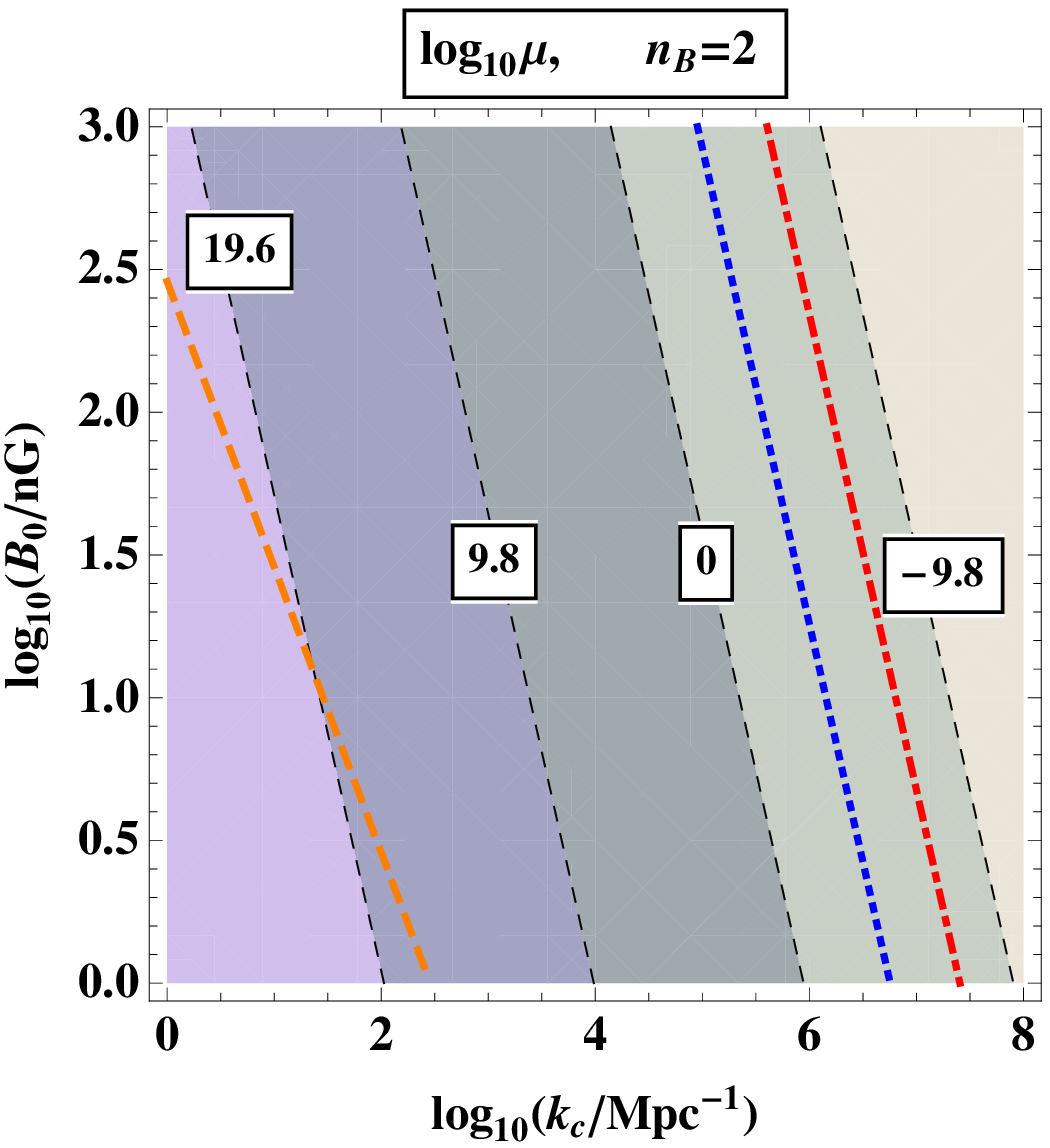}
\hspace{0.9cm}
\epsfxsize=2.9in\epsfbox{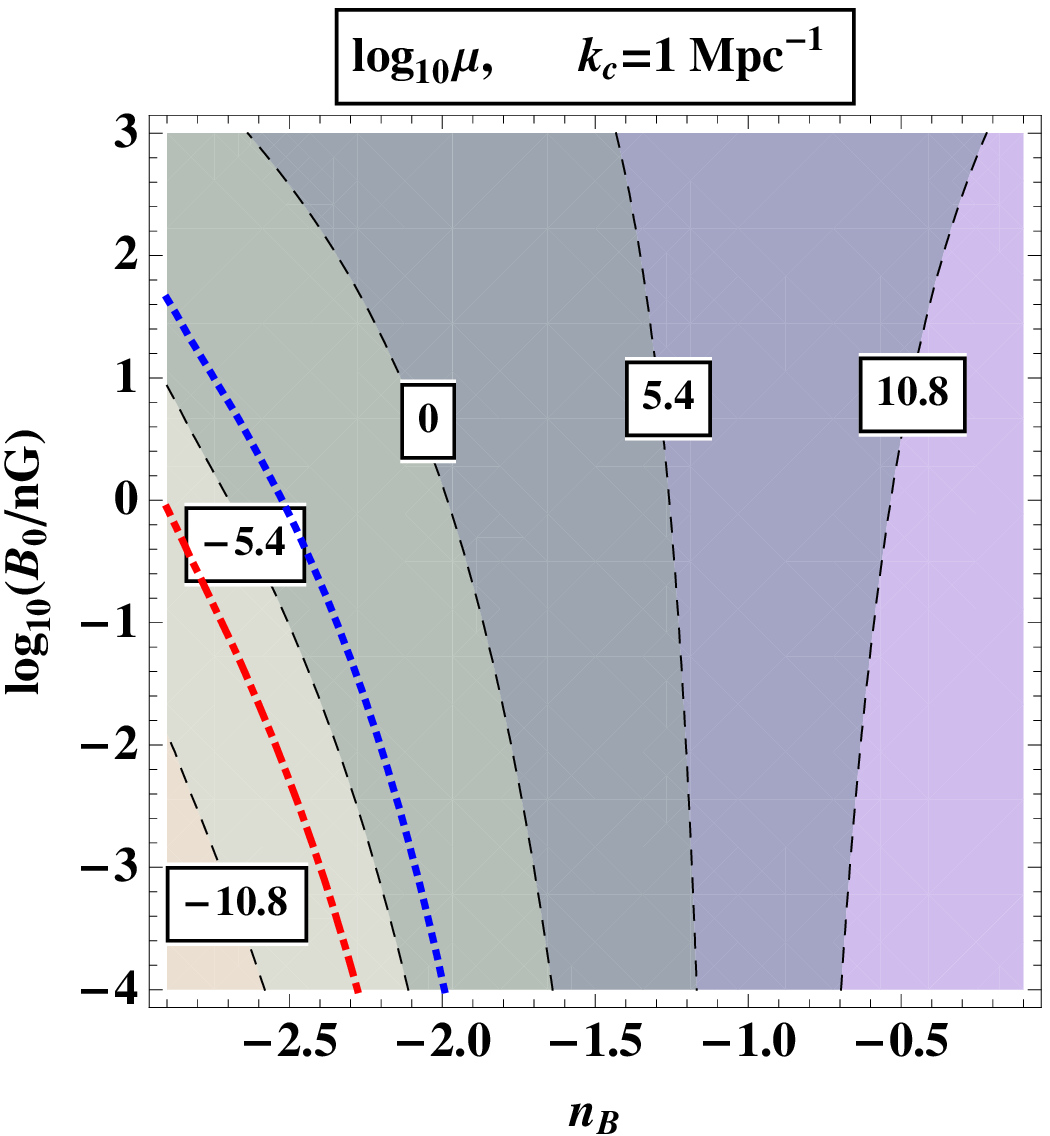}}
\caption{Contour plots of $\log_{10}\mu$ in the ($B_0$, $k_c)$ plane for
 a fixed index $n_B=2$ (left panel) and in the ($B_0$, $n_B)$ plane for
 a fixed smoothing scale $k_c=1~{\rm Mpc}^{-1}$ (right panel).
 The blue (dotted) lines show the COBE/FIRAS limit, $\mu=9.0\times
 10^{-5}$ \cite{firas3}, and the red (dot-dashed) lines show the 
projected PIXIE limit, $\mu=5.0\times 10^{-8}$ \cite{pixie}.
The orange (dashed) line in the left panel shows $k_c$ corresponding to
 the maximal value of the damping wavelength at decoupling
 (eq.~(\ref{kdast}) with 
 $\cos\theta=1$).}
\label{fig2}
\end{figure}
\begin{figure}
\centerline{\epsfxsize=2.9in\epsfbox{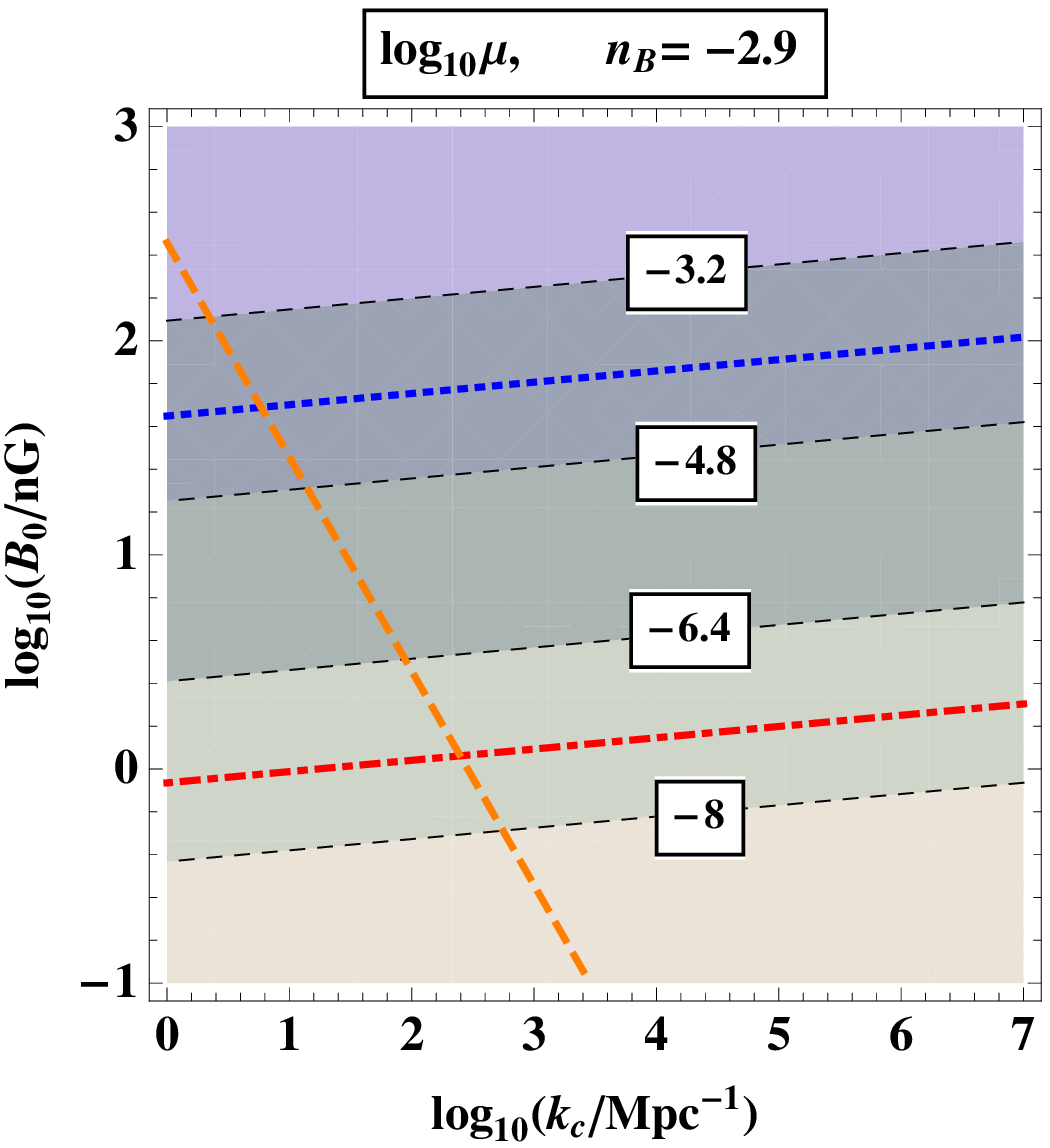}
\hspace{0.9cm}
\epsfxsize=2.9in\epsfbox{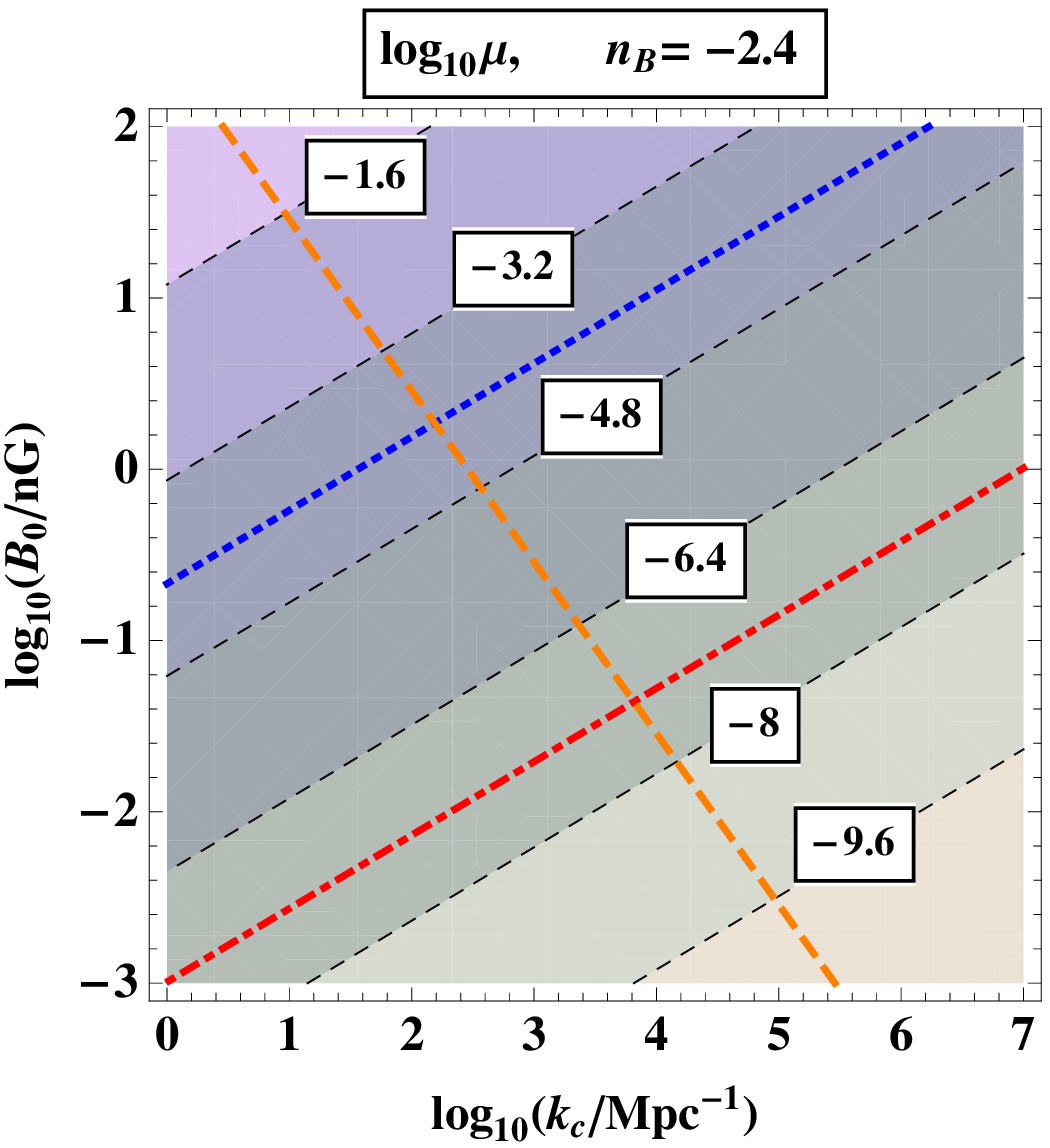}}
\caption{Same as the left panel of figure~\ref{fig2}, but for $n_B=-2.9$
 (left panel) and $-2.4$ (right panel).}
\label{fig4}
\end{figure}

The left panel of figure~\ref{fig2} shows how sensitive the inferred
values of $B_0$ are to the assumed values of $k_c$ for $n_B=2$. For
example, the projected PIXIE limit is satisfied for, say, $B_0=1$~nG, as
long as $k_c>3\times 10^7$ Mpc$^{-1}$, or a present-day
smoothing width of $<0.2$~pc. Figure~\ref{fig4} shows the same for
$n_B=-2.9$ (left panel) and $-2.4$ (right panel).

The right panel of figure~\ref{fig2} is more interesting. It  shows
which pairs of ($B_0$, $n_B$) are allowed for a given value of
$k_c=1~{\rm Mpc}^{-1}$.  We find
$B_0<10^{-4}$~nG for $n_B>-2.0$ from the COBE/FIRAS limit. On the other
hand, for $B_0=1$~nG, the COBE/FIRAS limit excludes $n_B>-2.6$. Both
limits can be improved significantly by PIXIE.

\begin{figure}
\centerline{\epsfxsize=2.9in\epsfbox{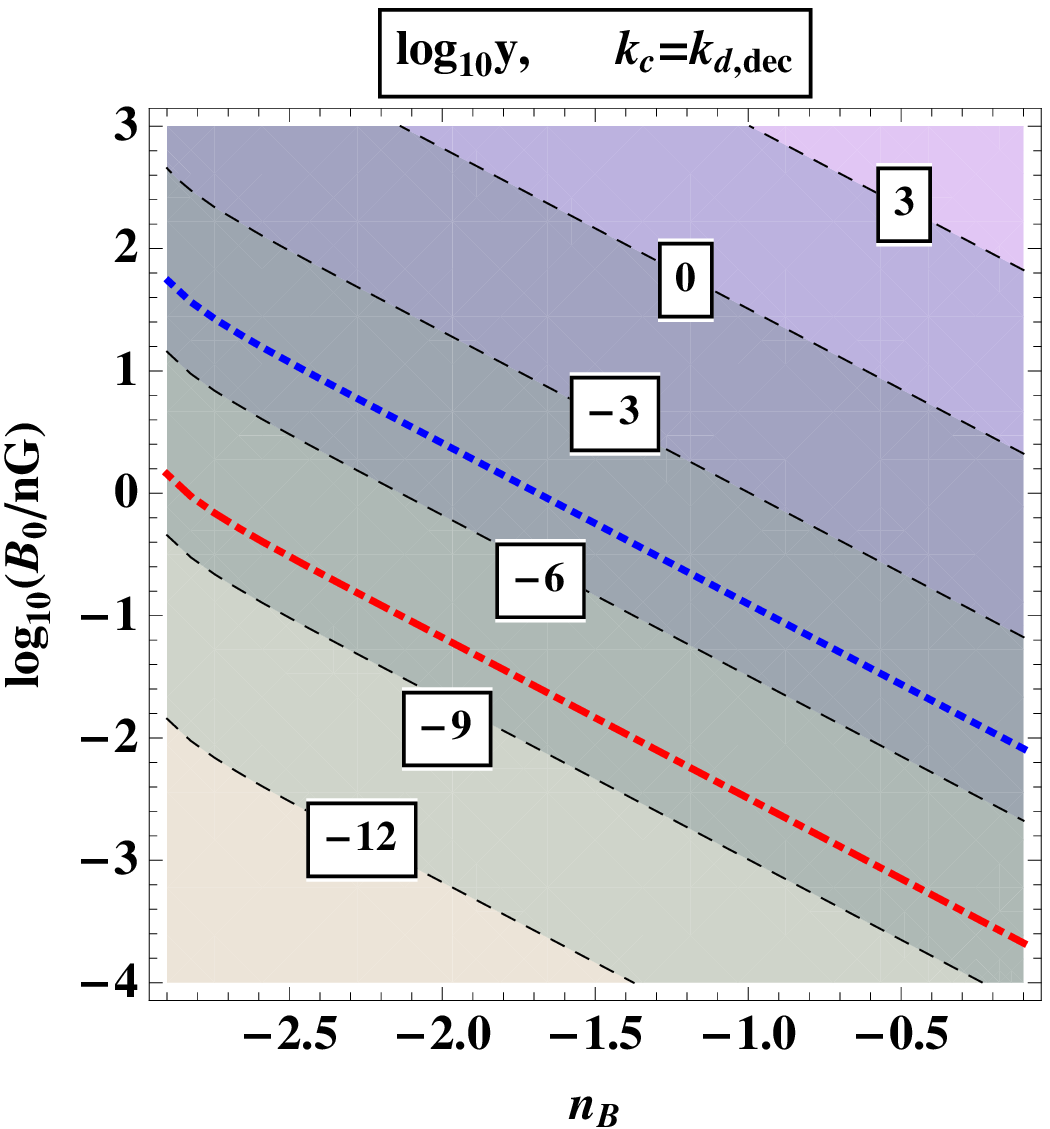}
\hspace{0.9cm}
\epsfxsize=2.9in\epsfbox{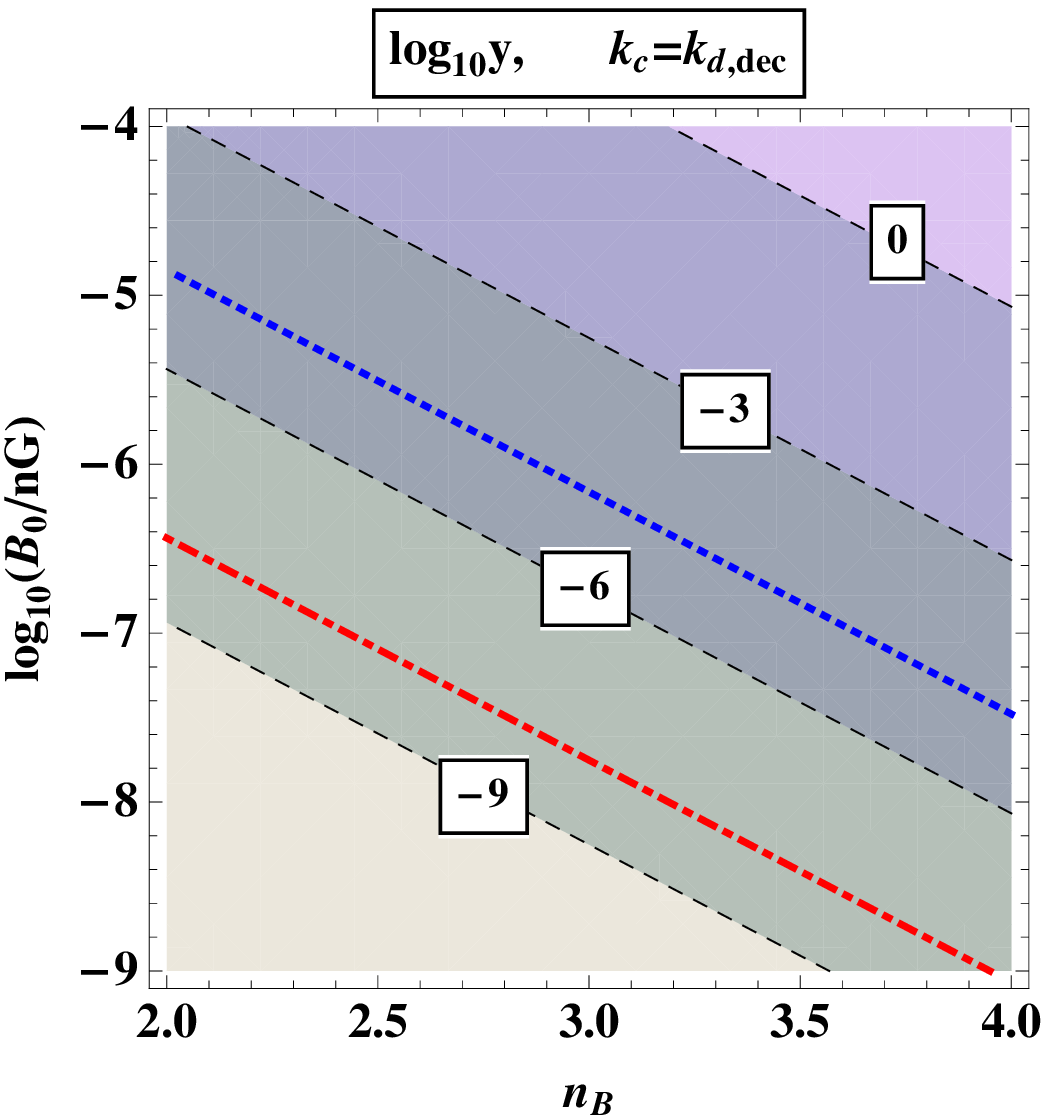}}
\caption{Same as figure~\ref{fig1} but for
 $\log_{10}y$.}
\label{fig5}
\end{figure}
\begin{figure}
\centerline{\epsfxsize=2.9in\epsfbox{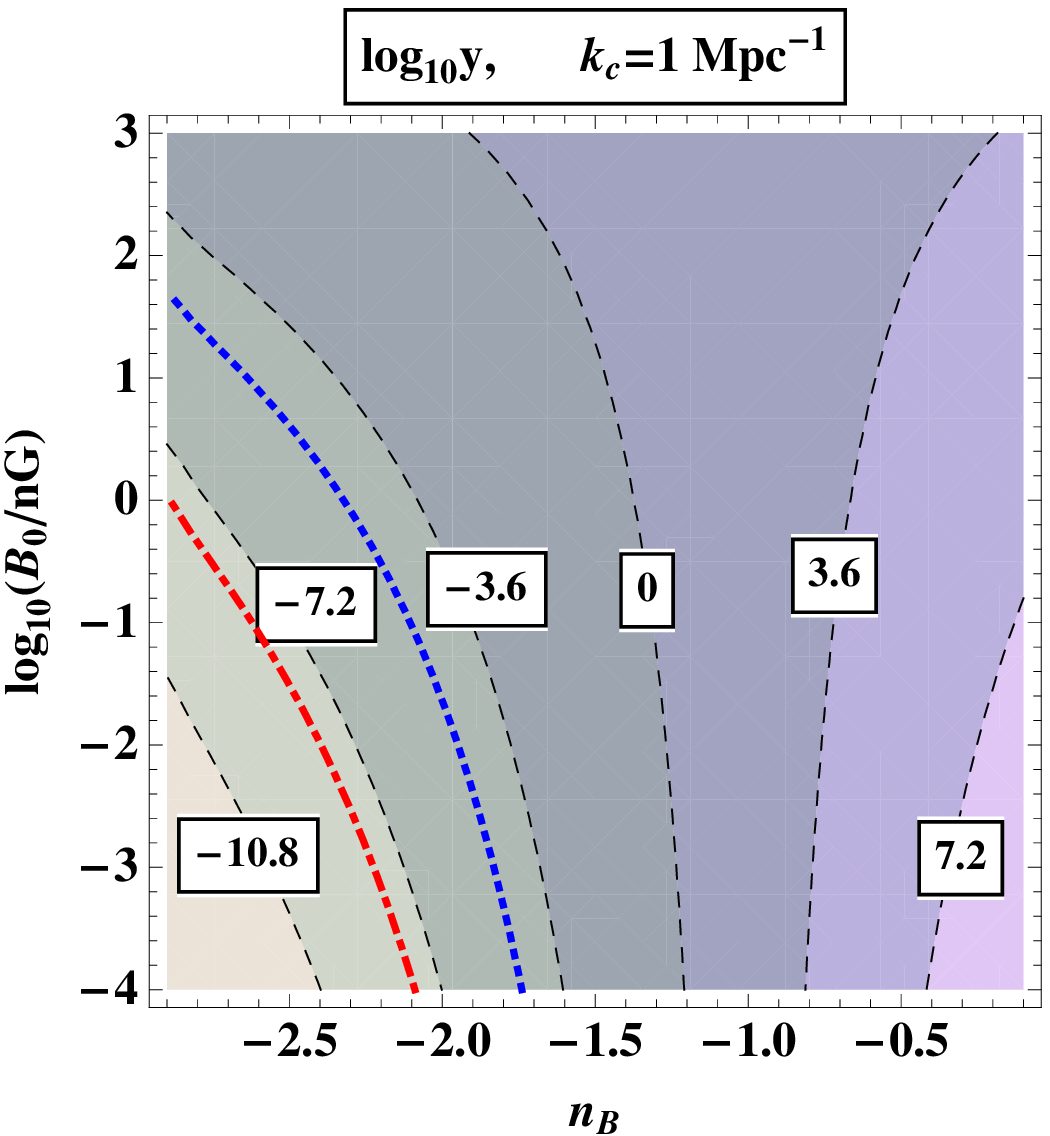}
\hspace{0.9cm}
\epsfxsize=2.9in\epsfbox{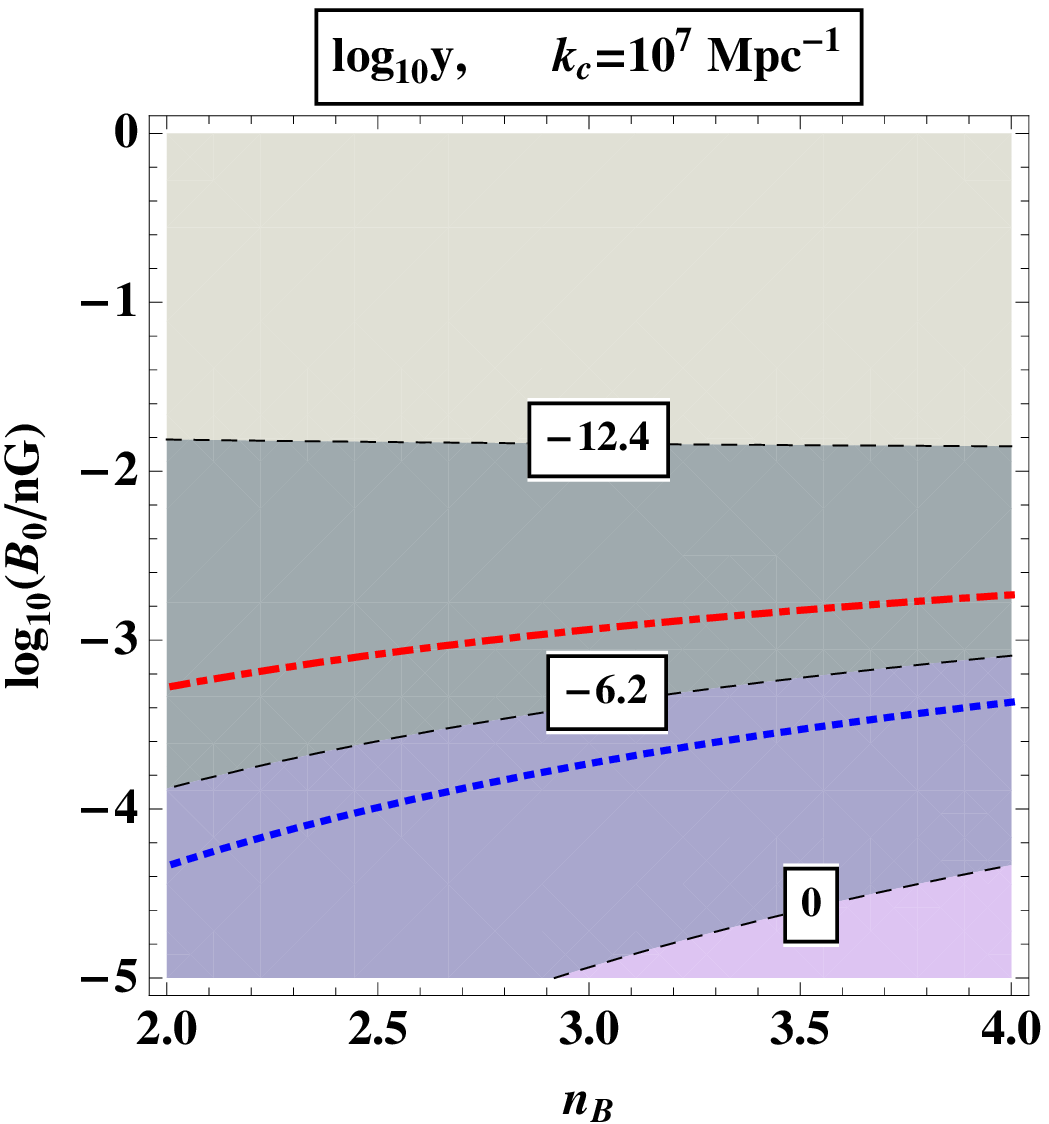}}
\caption{Same as the right panel of figure~\ref{fig2}, but for
 $\log_{10}y$ (left panel) and for $\log_{10}y$, $k_c=10^7~{\rm
 Mpc}^{-1}$, and $2\le n_B\le 4$ (right panel).}
\label{fig7}
\end{figure}
\begin{figure}
\centerline{\epsfxsize=2.9in\epsfbox{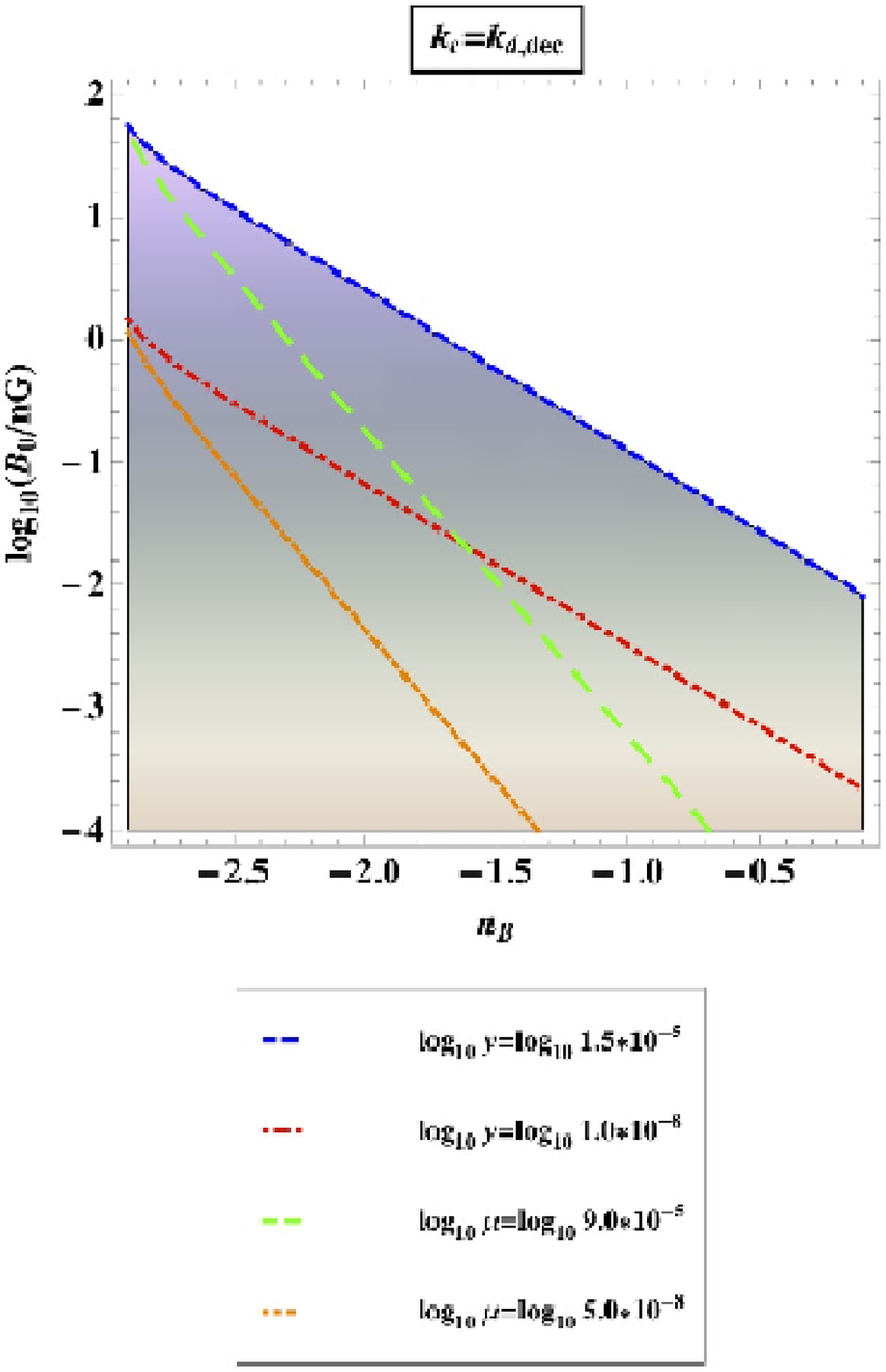}
\hspace{0.9cm}
\epsfxsize=2.9in\epsfbox{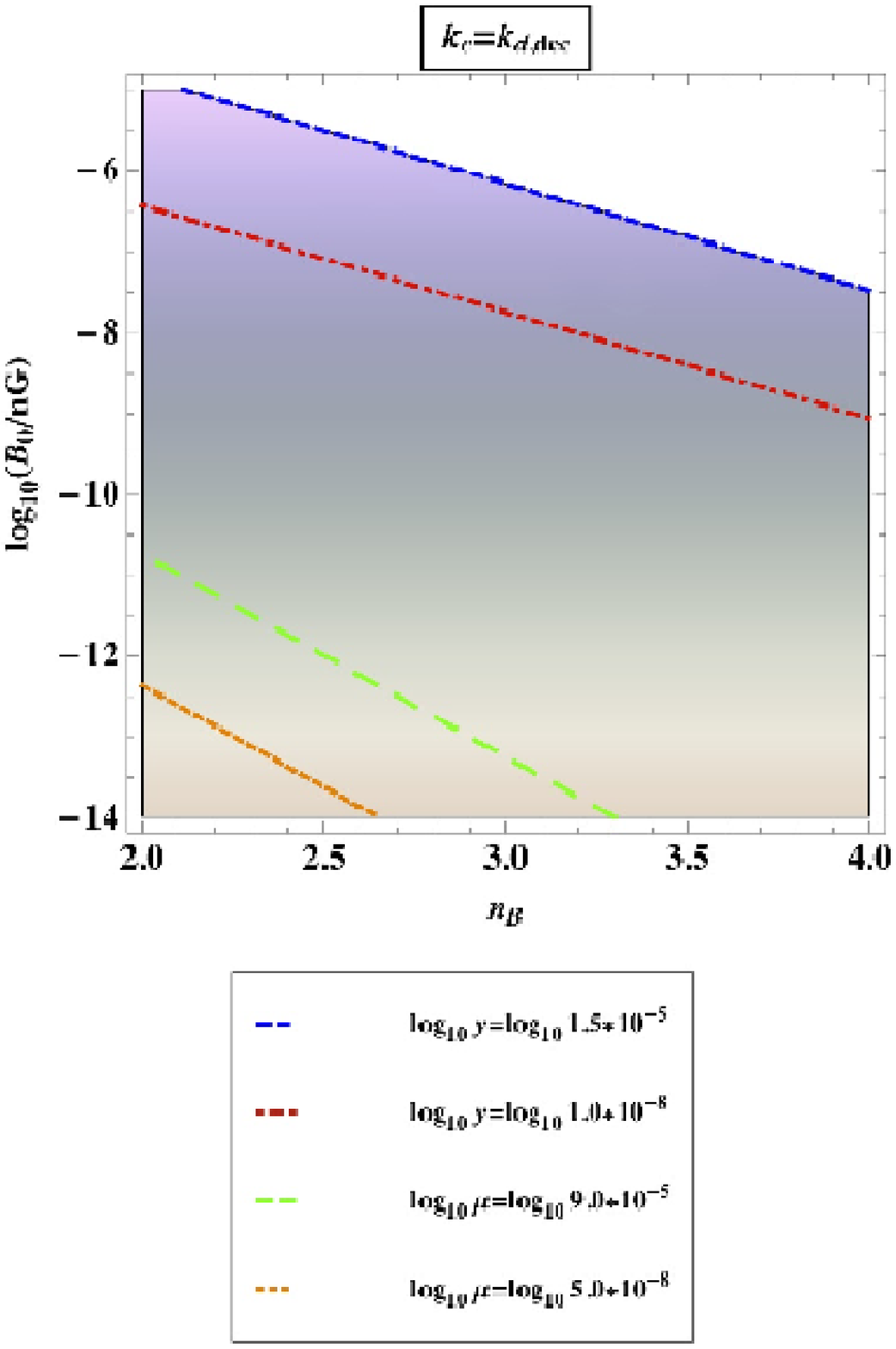}}
\caption{Comparison of constraints on $B_0$ and $n_B$ from 
 pre-decoupling $\mu$- and $y$-type distortions.
 The left and right panels show  $n_B$ expected from
 inflation and phase transitions, respectively.
The smoothing scale is set to be the maximal 
 value of the damping wavelength at decoupling (eq.~(\ref{kdast}) with
 $\cos\theta=1$).   
The blue (short-dashed) and green (long-dashed) lines show the
 COBE/FIRAS limits, $y=1.5\times 10^{-5}$ and $\mu=9.0\times
 10^{-5}$ \cite{firas3}, respectively. 
The red (dot-dashed) and orange (dotted) lines show the 
projected PIXIE limits, $y=10^{-8}$ and $\mu=5.0\times 10^{-8}$
 \cite{pixie}, respectively.}
\label{fig6}
\end{figure}
\begin{figure}
\centerline{\epsfxsize=2.9in\epsfbox{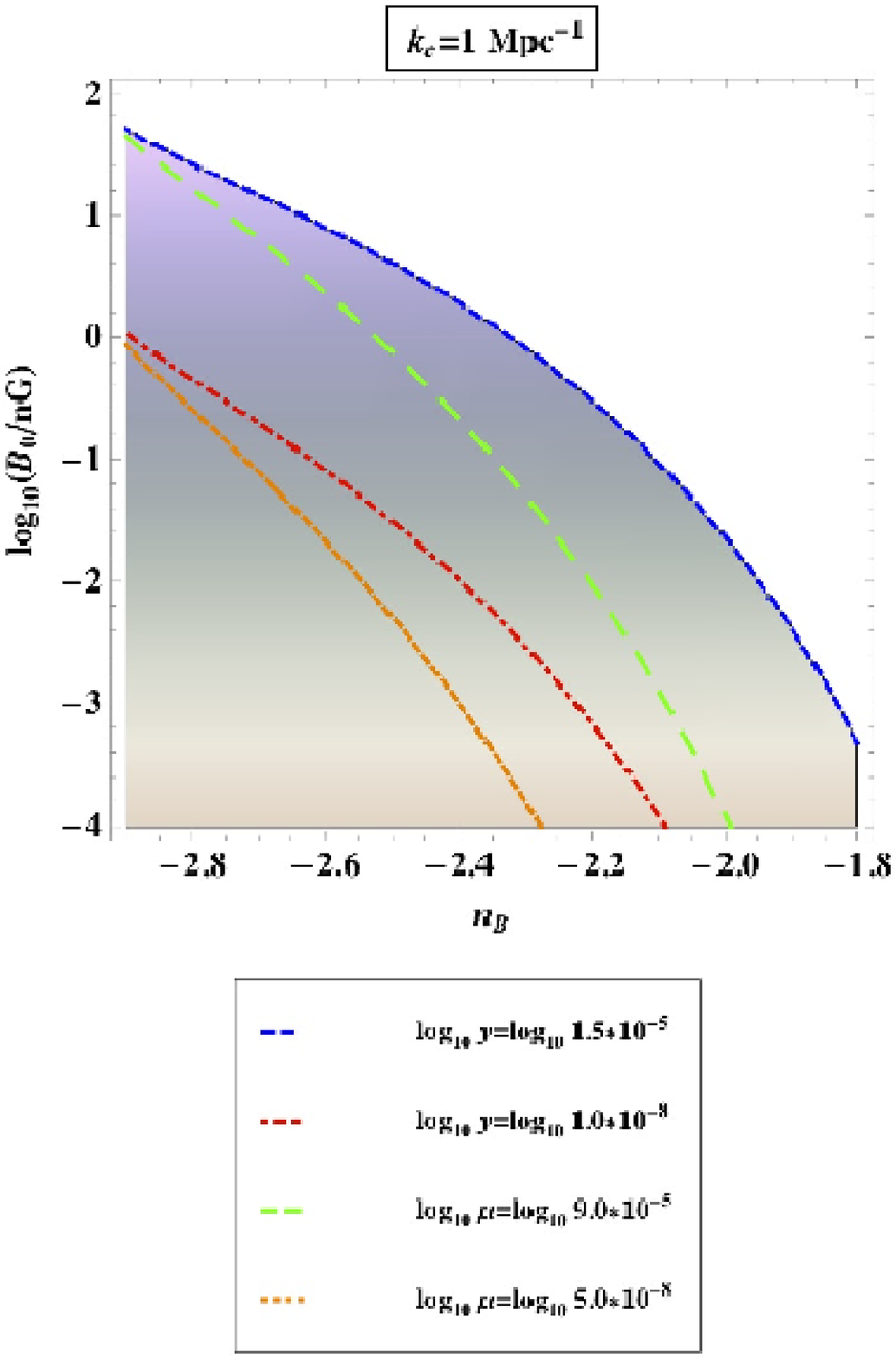}
\hspace{0.9cm}
\epsfxsize=2.9in\epsfbox{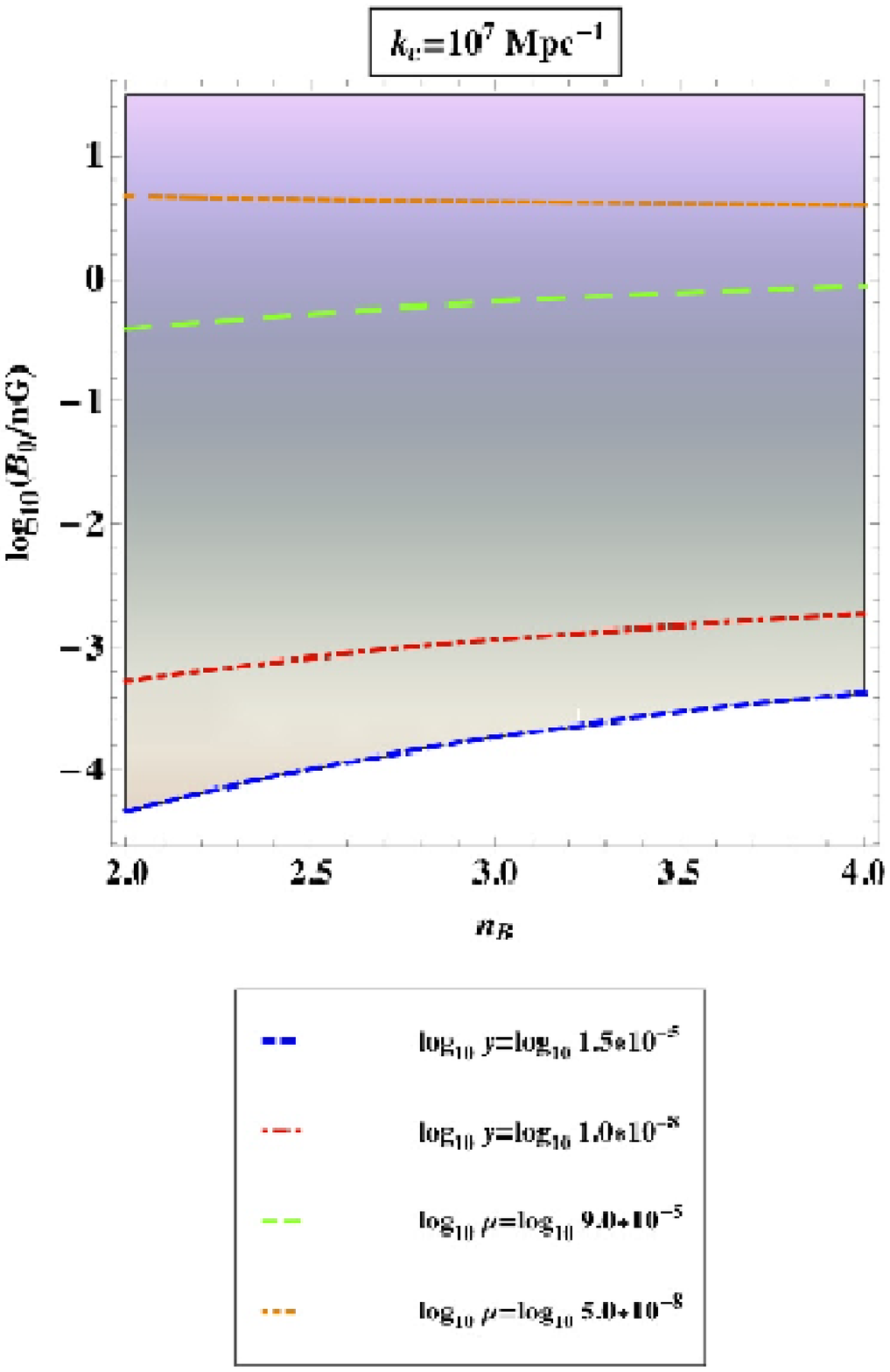}}
\caption{Comparison of constraints on $B_0$ and $n_B$ from 
 pre-decoupling $\mu$- and $y$-type distortions. (Left panel) Same as the left
 panel of figure~\ref{fig6}, but for $k_c=1~{\rm Mpc}^{-1}$ and $-2.9\le
 n_B\le -1.8$. (Right panel) Same as the right panel of
 figure~\ref{fig6}, but for $k_c=10^7~{\rm Mpc}^{-1}$.}
\label{fig8}
\end{figure}

\subsubsection{$y$-type distortion}

Next,  we calculate the $y$-type distortion due to dissipation of slow
magnetosonic and Alfv\'en waves using eq.~(\ref{y}). 
Comparing 
figures~\ref{fig5} and \ref{fig1} (also see figure~\ref{fig6}), 
and the left panel of figure~\ref{fig7} and the right panel of
figure~\ref{fig2} (also see figure~\ref{fig8}),  
we find
that constraints on 
the parameters of primordial magnetic fields from the $y$-type
distortion is much weaker than those from the $\mu$-type distortion for
non-scale-invariant spectra, $n_B>-3$. The reason is simple: as the
$y$-type distortion is created by energy injection at lower redshifts,
$z\lesssim 5\times 10^4$, the damping
wavenumber for the $y$-type distortion is much smaller than that for the
$\mu$-type distortion. As a result, the dissipated magnetic energy
creating the $y$-type distortion is much smaller than that for the
$\mu$-distortion for non-scale-invariant spectra, $n_B>-3$. For a
scale-invariant spectrum, such a change in the damping wavenumber does
not affect the result much, and thus $y$- and $\mu$-type
distortions are equally powerful for constraining the parameters of
primordial magnetic fields.

\section{Post-decoupling era}
\label{sec:postdecoupling}
After the decoupling epoch, there are two processes by which the
magnetic field can dissipate energy into the intergalactic medium. 
Firstly, there is ambipolar diffusion which arises due to the existence
of a remnant ionized component in the nearly completely neutral plasma
after decoupling. Secondly, non-linear effects can lead to
decaying MHD turbulence. In this section, we shall follow
Sethi and Subramanian \cite{sesu} to compute heating of the
intergalactic medium (IGM) 
due to dissipation of fields via ambipolar diffusion and decaying MHD
turbulence, and then compute the $y$-type distortion from inverse Compton
scattering of CMB photons off those heated electrons in the IGM.

\subsection{Ambipolar diffusion}
Ambipolar diffusion has its origin in the velocity difference between
the neutral and ionized components of matter due to the Lorentz force
only acting on the ionized component. 
The ionized components that are accelerated by the Lorentz force share their kinetic energy
with the neutral components via ion-neutral collisions, heating the
IGM. Therefore, the energy in magnetic fields is dissipated into the
IGM, as the velocity difference between ionized and neutral components
is damped by the collisions. 

The volume rate of energy dissipation due to ambipolar diffusion is
proportional to the average Lorentz force squared, and is given by \cite{sesu}
\begin{eqnarray}
\Gamma_{\rm
 in}=\frac{\rho_n}{16\pi^2\gamma\rho_b^2\rho_i}|(\vec{\nabla}\times\vec{B})\times\vec{B}|^2,
\label{eq:gammain}
\end{eqnarray}
where $\rho_n$, $\rho_i$, and $\rho_b$ are the energy densities of
neutral hydrogen, ionized hydrogen, and the total baryons,
respectively, and $\gamma$ is  the coupling between the ionized and neutral
component given by $\gamma\simeq\langle\sigma v\rangle_{H^+,H}/(2m_H)$, where $\langle\sigma v\rangle_{H^+,H}=0.649~T^{0.375}\times
10^{-9}$~cm$^3$~s$^{-1}$ \cite{sbk2}. 

In the Lorentz force,
$\vec{L}=(\vec{\nabla}\times\vec{B})\times\vec{B}$, the
derivative operator is defined with respect to the proper
coordinates. The flux freezing thus gives a strong redshift evolution of the
Lorentz force, $\vec{L}\propto (1+z)^{5}$. Recalling
$\rho_n/\rho_i\propto (1-x_e)/x_e$ 
and 
$\rho_b\propto (1+z)^3$, we find that the energy dissipation rate via
ambipolar diffusion
evolves as $\Gamma_{\rm in}\propto (1+z)^{3.625}(1-x_e)/x_e$
when the matter temperature goes as $T\propto 1/a$, and $\propto
(1+z)^{3.25}(1-x_e)/x_e$ when $T\propto 1/a^2$.
On the other hand,
as we shall show later, the energy dissipation rate via decaying MHD
turbulence evolves as $\Gamma_{\rm decay}\propto (1+z)^{11/2}$ (see eq.~(\ref{eq:gammadecay})); thus,
the ambipolar diffusion dominates at lower redshifts.

If velocities can still be treated as linear perturbations, the baryon
velocity in the standard $\Lambda$CDM model is a scalar
mode. Therefore, the velocity differences between the ionized and
neutral components are determined by the corresponding scalar mode of
the Lorentz force (e.g., \cite{kk1}):
\begin{eqnarray}
L_i=\frac{\rho_{\gamma}}{3a}\sum_{\vec{k}}kL(\vec{k})Y_i(\vec{k},\vec{x}),
\end{eqnarray}
where $Y_i=-k^{-1}Y_{|i}$. In flat space, the scalar harmonics have the
representation of $Y(\vec{k},\vec{x})=e^{i\vec{k}\cdot\vec{x}}$. 
The power spectrum of magnetic fields can be used to find the
corresponding power spectrum of $L(\vec{k})$ defined by 
\begin{eqnarray}
\langle L^*(\vec{k})L(\vec{q})\rangle=(2\pi)^3\delta(\vec{k}-\vec{q})P_L(k).
\end{eqnarray}
In Ref.~\cite{kk1}, $P_L(k)$ has been calculated using a Gaussian
damping with a width of $k_d^{-1}$ as in section~\ref{sec:predecoupling}.
The average Lorentz force is then given by
\begin{eqnarray}
\langle\vec{L}^2(\vec{r},z)\rangle&=&\frac{k_d^2\rho_{B,0}^2}{\left[\Gamma\left(\frac{n_B+3}{2}\right)\right]^2}(1+z)^{10}
\int_0^{\infty}dw~ w^{2n_B+7}e^{-w^2}\int_0^{\infty}dv~ v^{n_B+2}e^{-2v^2w^2}\int_{-1}^1dx~ e^{2w^2vx}
\nonumber\\
&&\times\left(1-2vx+v^2\right)^{\frac{n_B-2}{2}}
\left[1+2v^2+(1-4v^2)x^2-4vx^3+4v^2x^4\right],
\end{eqnarray}
where $w=k/k_d$, $v=q/k$, and $x=\frac{\vec{k}\cdot\vec{q}}{kq}$.
This can be written in terms of $B(z)$ and $l_d=k_d^{-1}/(1+z)$, so that
\begin{eqnarray}
\langle\vec{L}^2(\vec{r},z)\rangle&=&\left(\frac{B^4(z)}{l_d^2(z)}\right)
\frac{1}{4\left[\Gamma\left(\frac{n_B+3}{2}\right)\right]^2}
\int_0^{\infty}dw~ w^{2n_B+7}e^{-w^2}\int_0^{\infty}dv~ v^{n_B+2}e^{-2v^2w^2}\int_{-1}^1dx~ e^{2w^2vx}
\nonumber\\
&&\times\left(1-2vx+v^2\right)^{\frac{n_B-2}{2}}
\left[1+2v^2+(1-4v^2)x^2-4vx^3+4v^2x^4\right].
\label{avL}
\end{eqnarray}

\begin{figure}
\centerline{\epsfxsize=2.2in\epsfbox{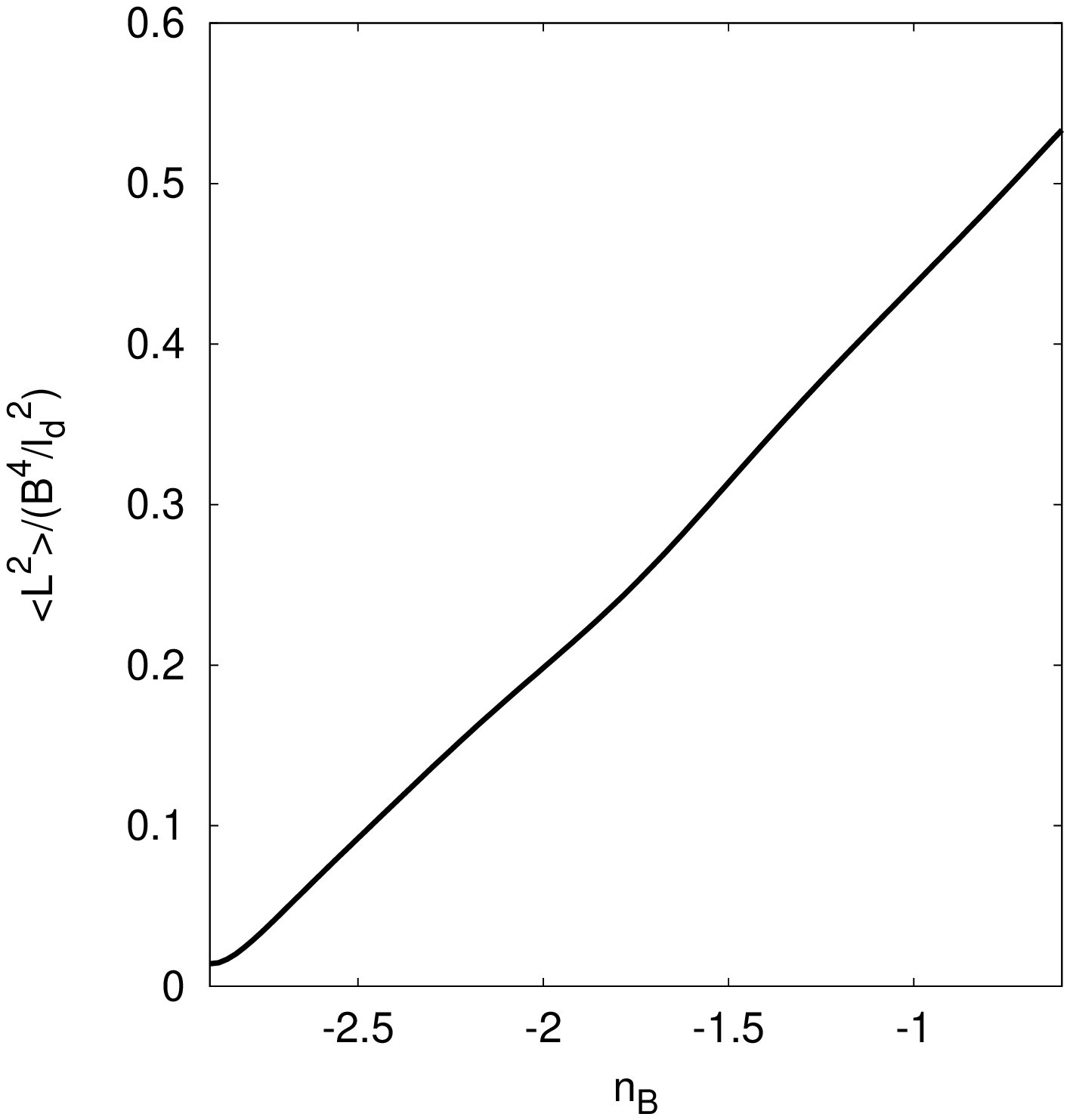}
\hspace{2.2cm}
\epsfxsize=2.2in\epsfbox{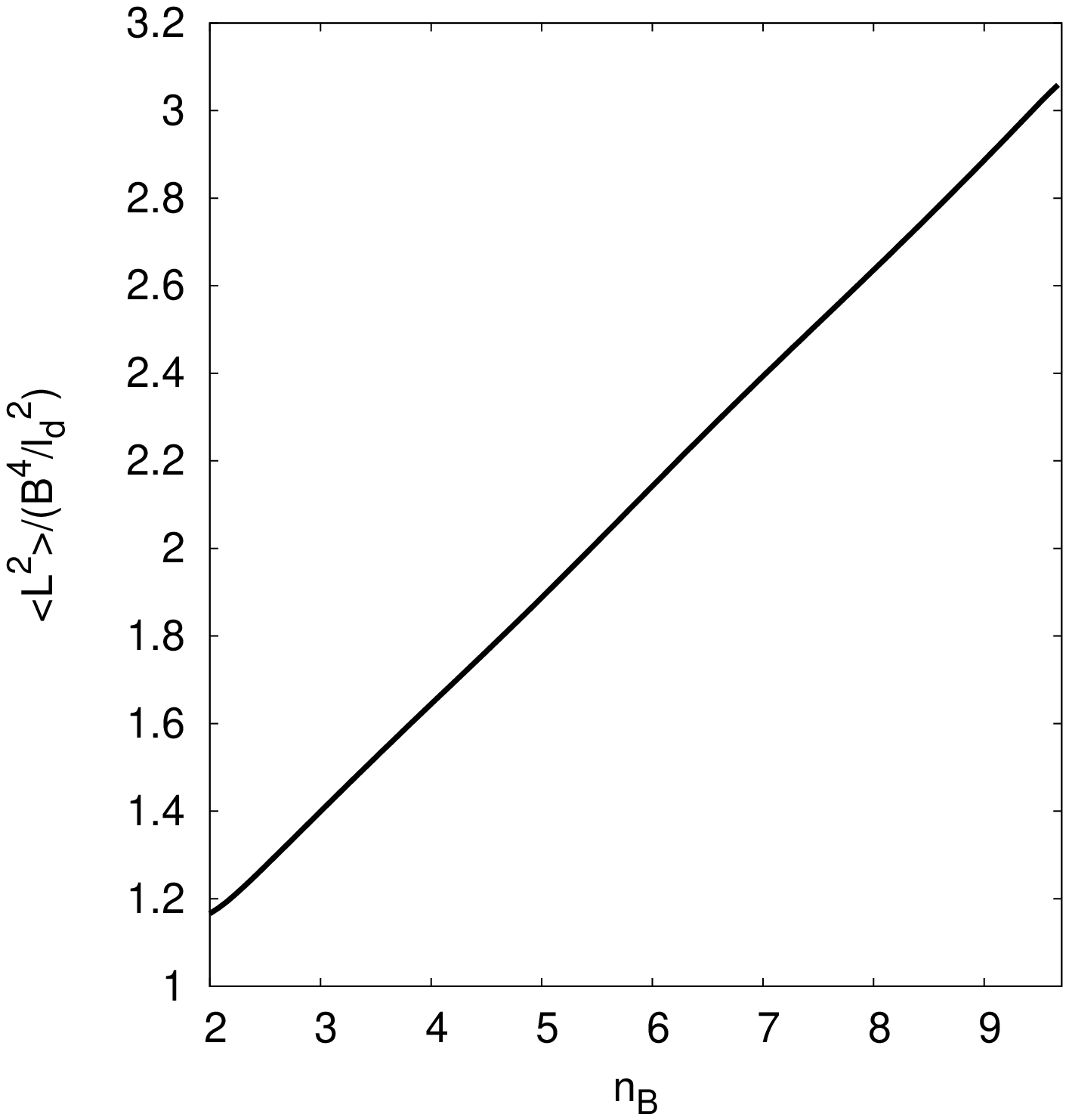}}
\caption{Ratio of the average Lorentz term calculated from
 eq.~(\ref{avL}) and  $B^4/l_d^2$ as a function of
 $n_B$. }
\label{fig9}
\end{figure}

In Refs.~\cite{sbk1,sbk2} the average Lorentz term has been approximated
by $B^4/l_d^2$. In figure~\ref{fig9}, we show the ratio of the average
Lorentz term calculated from eq.~(\ref{avL}) and $B^4/l_d^2$. Clearly
$B^4/l_d^2$ is not a good approximation, and thus we shall use
eq.~(\ref{avL}) unless indicated otherwise. Moreover, $k_d$ is chosen to
be $k_{d,dec}$ given in eq. (\ref{kdast}) with $\cos\theta=1$.

\subsection{Decaying MHD turbulence}
Turbulent motion in the plasma is suppressed before the decoupling epoch
due to a large radiative viscosity. However, it is no longer suppressed
after the decoupling epoch, and the Reynolds number becomes very 
large. On scales smaller than the magnetic Jeans length, non-linear mode
interactions transfer energy to smaller scales, dissipating the magnetic
field on larger scales and inducing MHD turbulence
to decay.

One can use results of numerical simulations of MHD turbulence in flat
space to calculate the evolution of magnetic fields in an expanding
universe, by rescaling variables \cite{sim1,sim2}. The estimated decay
rate for a non-helical field in the matter-dominated era is
given by \cite{sesu}  
\begin{eqnarray}
\Gamma_{\rm decay}=\frac{B_0^2}{8\pi}\frac{3m}{2}\frac{\left[\ln\left(1+\frac{t_d}{t_i}\right)\right]^m}
{\left[\ln\left(1+\frac{t_d}{t_i}\right)+\ln\left[\left(\frac{1+z_i}{1+z}\right)^{\frac{3}{2}}\right]\right]^{m+1}}
H(t)(1+z)^4,
\label{eq:gammadecay}
\end{eqnarray}
where $B_0$ is again the present-day field value assuming a flux
freezing; $m$ is related to the magnetic spectral index as
$m=\frac{2(n_B+3)}{n_B+5}$; $t_d$ is the physical decay time scale for
turbulence given by $t_d/t_i=(k_J/k_d)^{\frac{(n_B+5)}{2}}
\simeq 14.8~(B_0/1~{\rm nG})^{-1}(k_d/1~{\rm
Mpc}^{-1})^{-1}$ 
with the magnetic Jeans wavenumber of $k_J\simeq
14.8^{\frac{2}{n_B+5}} (B_{0}/{\rm
1~nG})^{-\frac{2}{n_B+5}}\left(k_d/1~{\rm
Mpc}^{-1}\right)^{\frac{n_B+3}{n_B+5}}$ Mpc$^{-1}$ \cite{sesu}; and
$z_i$ and $t_i$ are the 
redshift and time at which dissipation of the magnetic field due to
decaying MHD turbulence becomes important. We use $z_i=z_{\rm dec}=1088$
\cite{wmap9}.  

Ignoring a logarithmic dependence on $1+z$, we find that the
energy dissipation rate via decaying MHD turbulence evolves as
$\Gamma_{\rm decay}\propto (1+z)^{11/2}$, which is faster than that of
ambipolar diffusion (see eq.~(\ref{eq:gammain})). Therefore, we expect
dissipation via decaying turbulence to dominate in early times.

\subsection{Results}
\subsubsection{Evolution of electron temperature and ionization fraction}

The evolution of the electron temperature is determined by \cite{sesu}
\begin{eqnarray}
\dot{T}_e=-2\frac{\dot{a}}{a}T_e+\frac{x_e}{1+x_e}\frac{8\rho_{\gamma}\sigma_T}{3m_ec}\left(T_{\gamma}-T_e\right)+\frac{x_e\Gamma}{1.5 k_B n_e},
\end{eqnarray}
where $\Gamma=\Gamma_{\rm in}$ for ambipolar diffusion;
$\Gamma=\Gamma_{\rm decay}$ for energy dissipation due to decaying MHD
turbulence; and $\Gamma=\Gamma_{\rm in}+\Gamma_{\rm decay}$ when both are
included in the calculation.

\begin{figure}
\centerline{\epsfxsize=2.2in\epsfbox{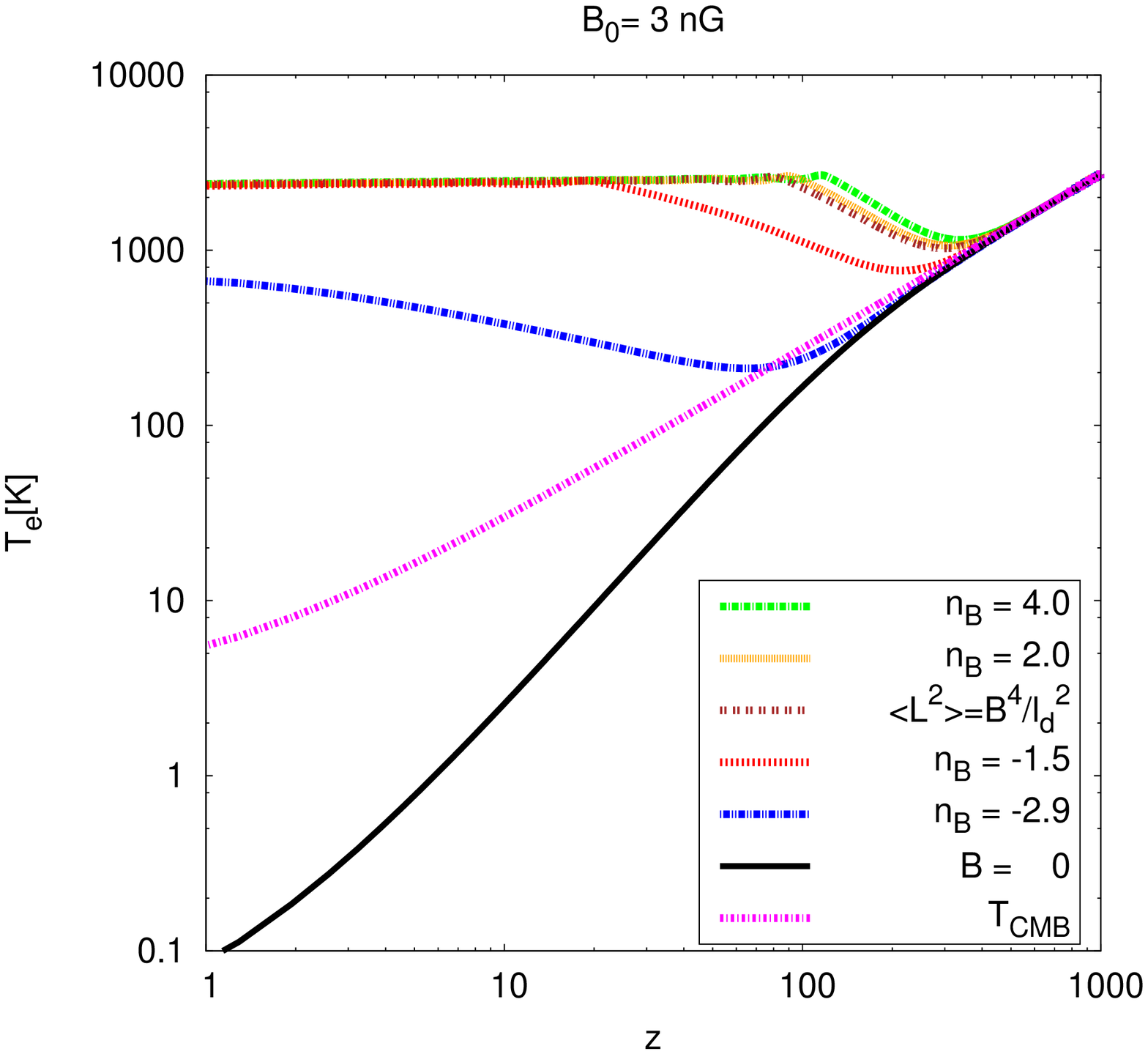}
\hspace{0.1cm}
\epsfxsize=2.2in\epsfbox{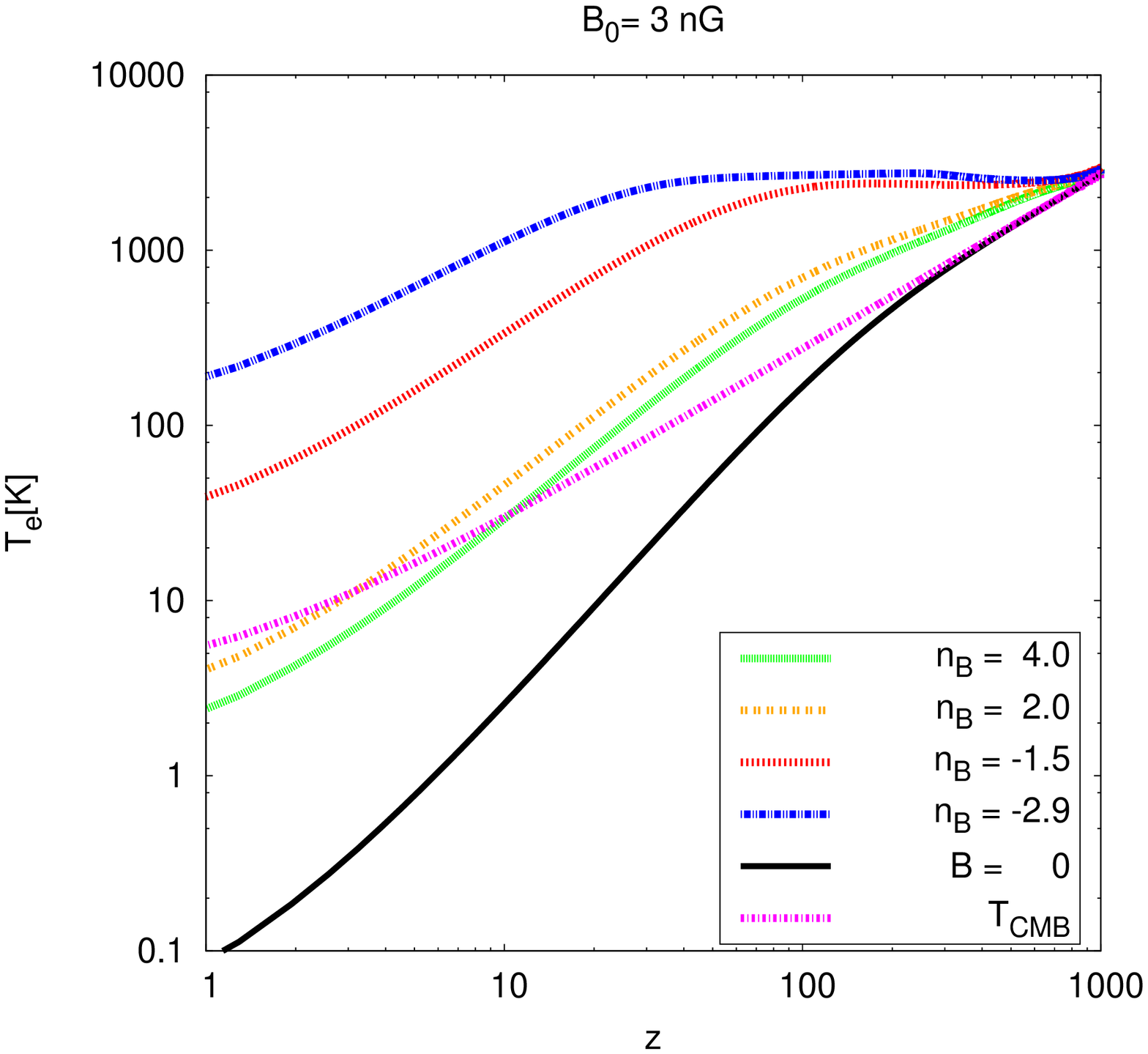}
\hspace{0.1cm}
\epsfxsize=2.2in\epsfbox{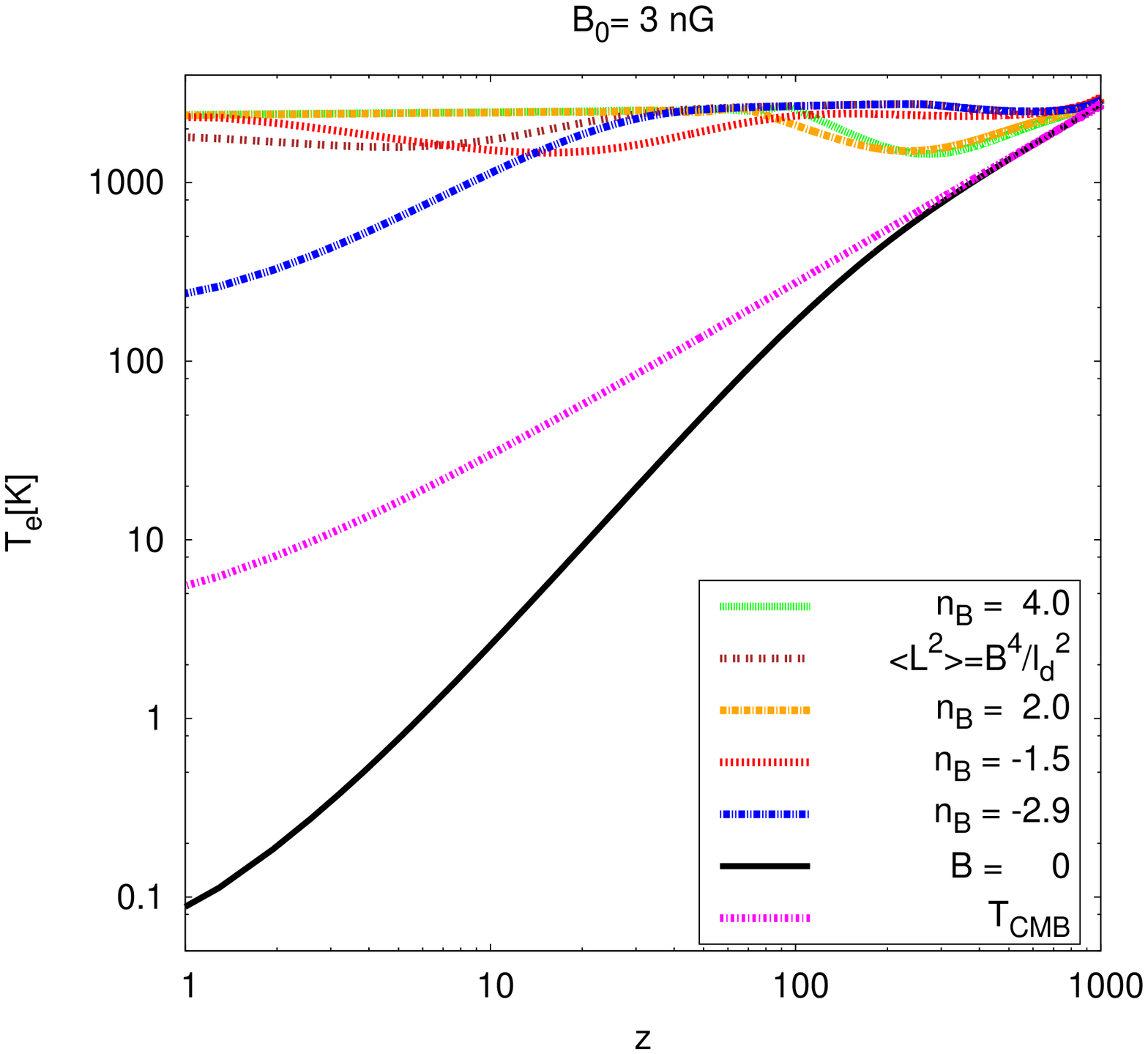}
}
\caption{Evolution of the matter temperature with and without
 dissipation of magnetic fields.
(Left panel) Dissipation by ambipolar
 diffusion. (Middle panel) Dissipation by decaying MHD turbulence. (Right 
 panel) Both contributions are included. For all cases, a present-day
 magnetic field of $B_0=3$~nG, smoothed over $k_{d,dec}$
 (eq.~(\ref{kdast}) with $\cos\theta=1$),
 is assumed.  We explore four different 
 values of the spectral index ($n_B=-2.9$, $-1.5$, $2$, and $4$). The
 black solid lines (the bottom line in each panel) do not include
 dissipation of magnetic fields. The magenta dotted lines (the second to
 the bottom line in 
 each panel) show the CMB temperature.
 The brown dotted lines in the left and right panels use the
 approximation for the squared Lorentz force, $\langle
 L^2\rangle=B^4/l_d^2$; otherwise we use eq.~(\ref{avL}).
}
\label{fig10}
\end{figure}

The evolution of the ionization fraction, $x_e$, is not directly
 affected by magnetic fields, but it is indirectly affected by changes
 in $T_e$ (i.e., higher $T_e$ gives more collisional
 ionization in the IGM).
The system of equations determining the thermal and ionization history
of the universe has been solved in the absence of magnetic fields
\cite{recfast1,recfast2,recfast3,recfast4,recfast5,recfast6,recfast7,recfast8,recfast9,recfast10}. We
have modified a public code {\tt RECFAST++}\footnote{{\tt
http://www.cita.utoronto.ca/$\tilde{\; }$jchluba/Science$\underline{
\;}$Jens/Recombination/Recfast++.html}}, which is the C version of
{\tt RECFAST}\footnote{{\tt
http://www.astro.ubc.ca/people/scott/recfast.html}} with some
improvements, by including  $\Gamma_{\rm in}$ and $\Gamma_{\rm decay}$ in the
evolution equation of $T_e$. 

Figure~\ref{fig10} shows the evolution of electron temperature with and
without dissipation of magnetic fields. The left and middle panels include
the contribution of either ambipolar diffusion or decaying MHD turbulence,
respectively, and the right panel includes both contributions. Compared
with non-magnetized case (the bottom black solid line in each panel), we
find that dissipation of magnetic fields raises the electron temperature
by many orders of magnitude. Ambipolar diffusion and decaying MHD
turbulence heat the gas at different epochs: the former is important 
at $z\lesssim 100$ while the latter is important at $z\gtrsim
100$. This is consistent with our expectation based on how
 the energy dissipate rate evolves: $\Gamma_{\rm in}\propto
 (1+z)^{3.625}(1-x_e)/x_e$ for $T\propto 1/a$ and
 $(1+z)^{3.25}(1-x_e)/x_e$ for $T\propto 1/a^2$,
 and $\Gamma_{\rm decay}\propto (1+z)^{5.5}$.

When both contributions are included, we find that the temperature is
raised at roughly all redshifts. For $B_0=3$~nG, we find that the
temperature is raised up to about 2600~K,
which is only a factor of about four  lower than a typical
temperature of photo-ionized gas ($10^4$~K).

We also find that the dependence of the temperature on $n_B$ is opposite
for ambipolar diffusion and decaying MHD turbulence. For ambipolar
diffusion, the larger the spectral index $n_B$ is, the higher the
temperature becomes, while the opposite happens for decaying MHD
turbulence. This is because of the scales at which dissipation
occurs. For ambipolar diffusion, small-scale fields are dissipated. 
For decaying MHD turbulence, energy in large-scale fields is transferred to
small scales and dissipates. 
Mathematically, $\Gamma_{\rm
in}$ is proportional to the Lorentz force squared, which is proportional
to $B^4$ with a factor of $k^2$ via spatial derivatives, whereas
$\Gamma_{\rm decay}$ 
is proportional to the energy density of fields, $B^2$. 

\begin{figure}
\centerline{\epsfxsize=2.2in\epsfbox{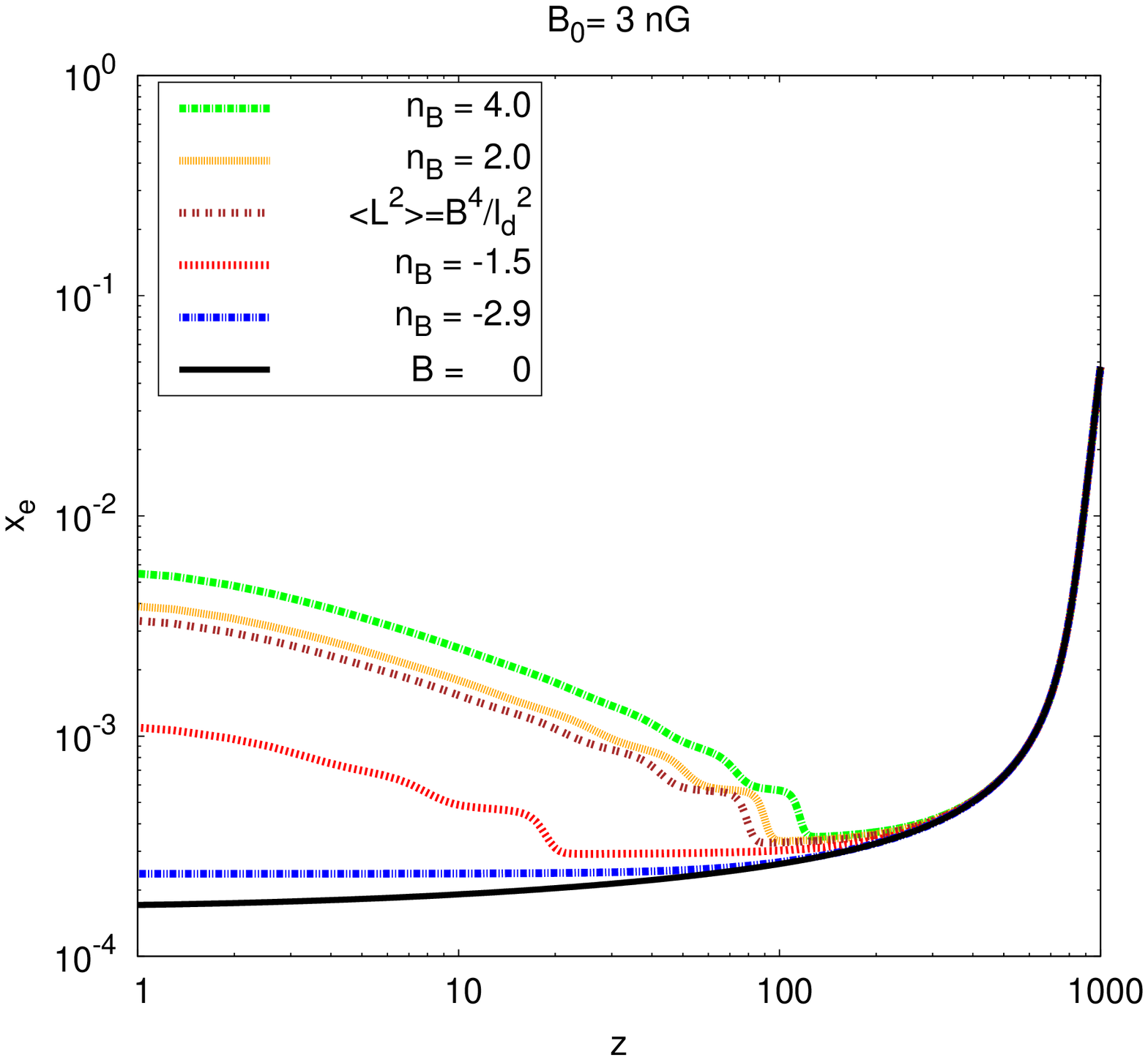}
\hspace{0.1cm}
\epsfxsize=2.2in\epsfbox{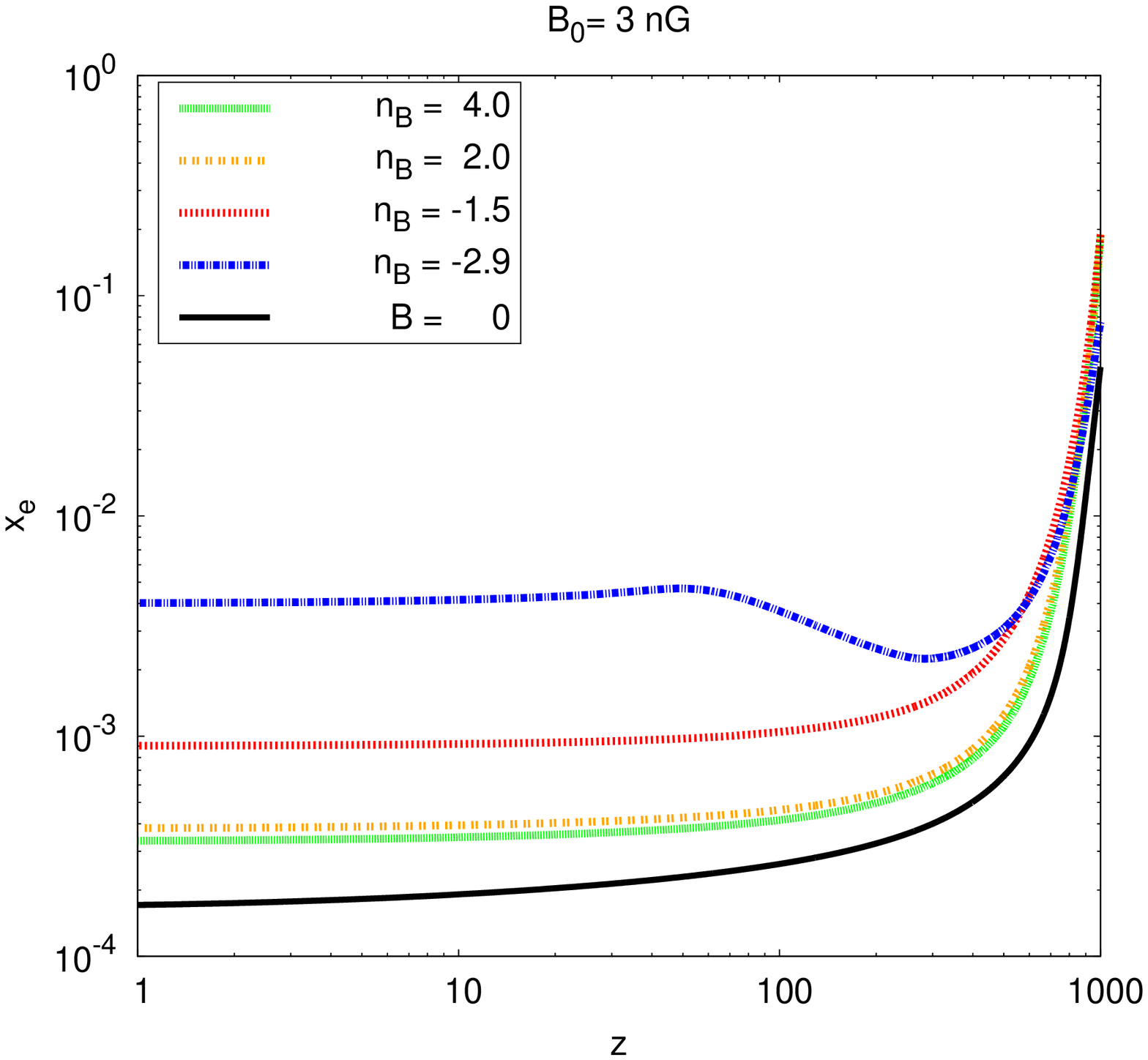}
\hspace{0.1cm}
\epsfxsize=2.2in\epsfbox{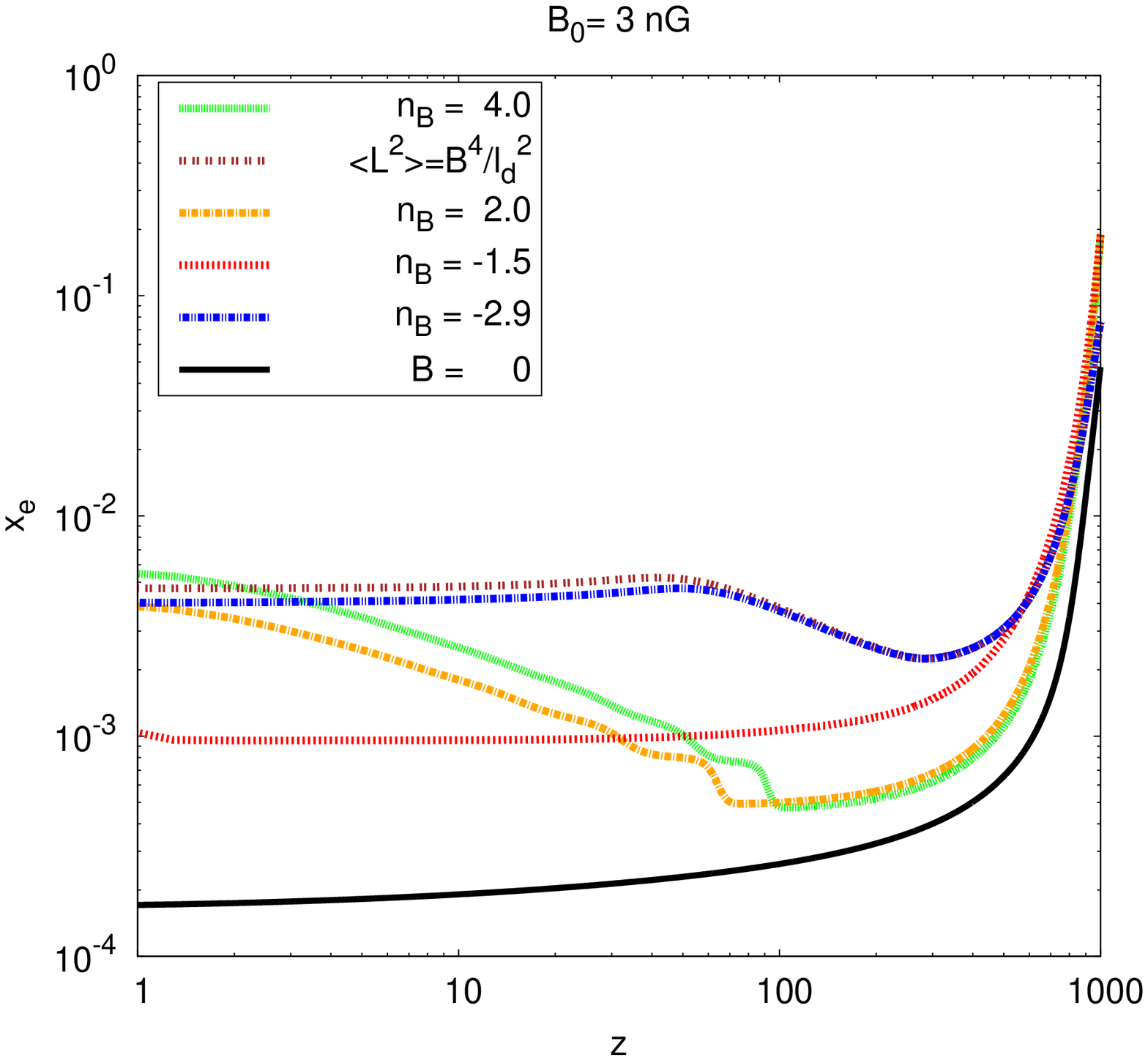}
}
\caption{Evolution of the ionization fraction, $x_e$, with and without
 dissipation of magnetic fields.
(Left panel) Dissipation by ambipolar
 diffusion. (Middle panel) Dissipation by decaying MHD turbulence. (Right 
 panel) Both contributions are included. For all cases, a present-day
 magnetic field of $B_0=3$~nG,  smoothed over $k_{d,dec}$
 (eq.~(\ref{kdast}) with $\cos\theta=1$), is assumed. We explore four different
 values of the spectral index ($n_B=-2.9$, $-1.5$, $2$, and $4$). The 
 black solid lines (the bottom line in each panel) do not include
 dissipation of magnetic fields. 
 The brown dotted lines in the left and right panels use the
 approximation for the squared Lorentz force, $\langle
 L^2\rangle=B^4/l_d^2$; otherwise we use eq.~(\ref{avL}).
}
\label{fig11}
\end{figure}

Figure~\ref{fig11} shows the evolution of ionization fraction,
$x_e$. For $B_0=3$~nG, the ionization fraction can reach of order
$10^{-3}$ or more, which is an order of magnitude greater than $x_e$
without dissipation of magnetic fields. The dependence of $x_e$ on $n_B$
is also similar to that of $T_e$. An important implication of this
calculation for the ionization history of the universe is that there can
be a significant, of order 0.1\%, ionization even at $z\approx 100$. 
Dissipation of magnetic fields plays only a sub-dominant role at
$z\lesssim 10$, in which the ionization fraction reaches unity by
reionization of the universe by first stars.

\begin{figure}
\centerline{\epsfxsize=2.9in\epsfbox{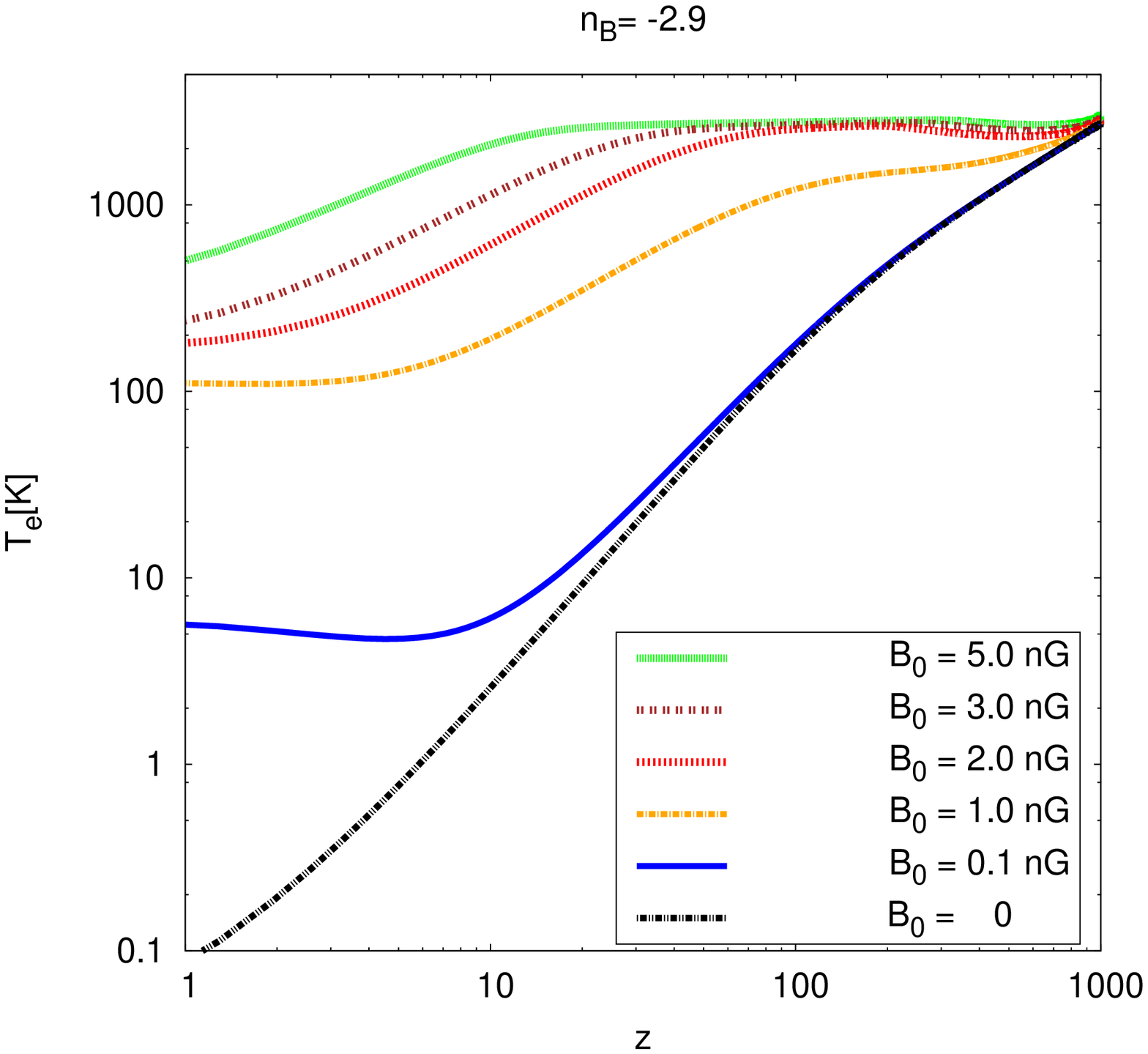}
\hspace{0.9cm}
\epsfxsize=2.9in\epsfbox{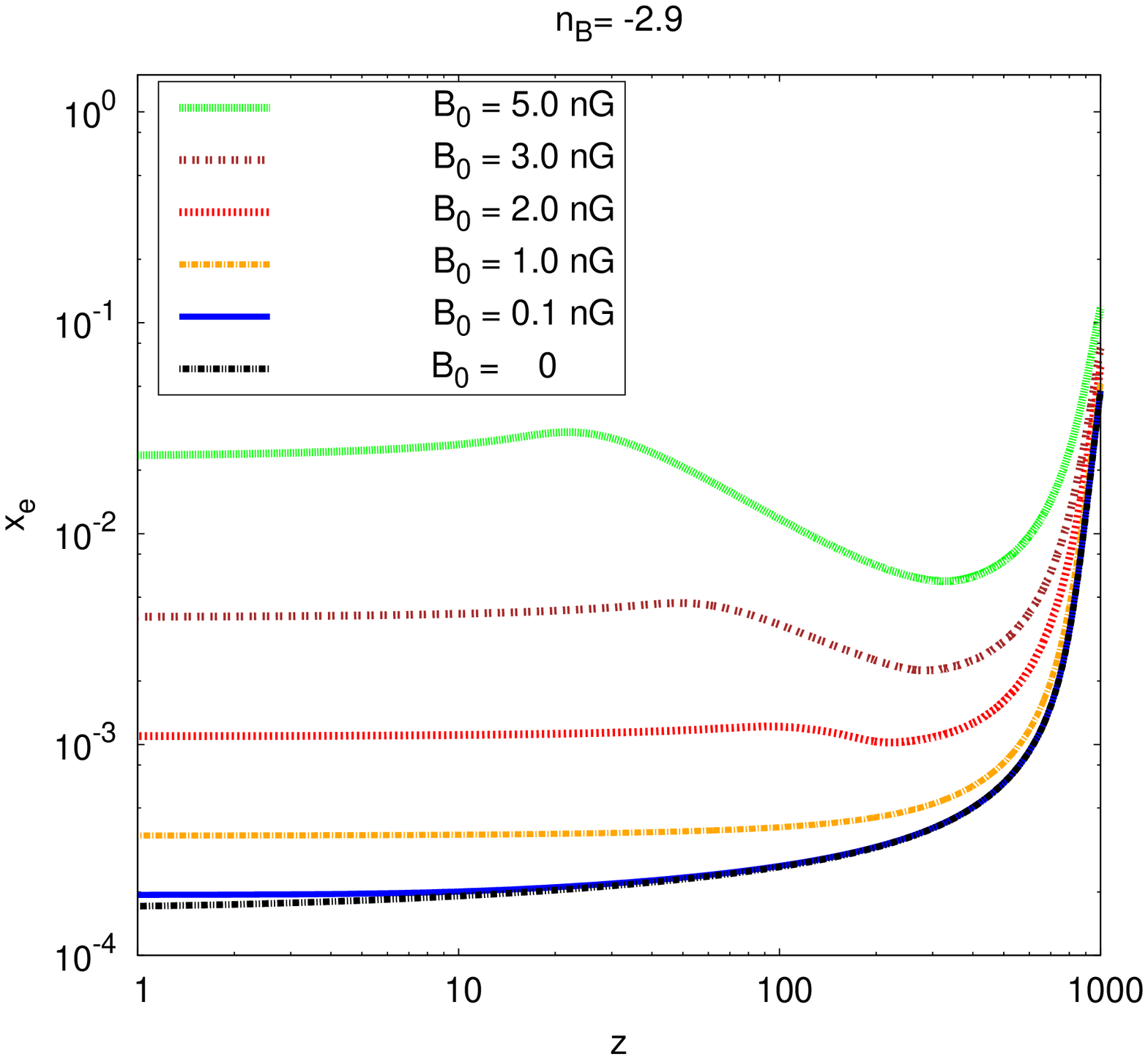}}
\caption{Evolution of the matter temperature (left panel) and the
 ionization fraction (right panel) with and without dissipation of
 magnetic fields. From top to bottom lines, the field values are
 $B_0=5$, 3, 2, 1, 0.1, and 0~nG, and the magnetic power spectrum is
 nearly scale invariant with $n_B=-2.9$. Both ambipolar diffusion and decaying
 MHD turbulence are included. 
}
\label{fig12}
\end{figure}

The rate of energy dissipation due to ambipolar diffusion is
proportional to $B^4_0$ (see eq.~(\ref{eq:gammain})) and that due to
decaying MHD turbulence is proportional to $B_0^2$ (up to logarithmic
dependence; see eq.~(\ref{eq:gammadecay})). Therefore, we would expect a
strong dependence of $T_e$ and $x_e$ on $B_0$. Figure~\ref{fig12} shows
that $T_e$ and $x_e$ depend sensitively on $B_0$; however, $T_e$ at
$z\gtrsim 100$ does not increase by going from 2~nG to 5~nG. 
This is because the plasma cools efficiently by the inverse
Compton scattering of the CMB photons off electrons at high
redshifts, and thus the temperature cannot rise indefinitely.

\subsubsection{$y$-type distortion}
We calculate the $y$-type distortion using
\begin{eqnarray}
y=-\int_{z_i}^{z_0}\frac{dz}{(1+z)}\frac{n_e\sigma_Tc}{H(z)}\frac{k_B(T_e-T_{CMB})}{m_ec^2}, 
\label{ypost}
\end{eqnarray}
where $z_i=z_{\rm dec}=1088$ \cite{wmap9} and $z_0=10^{-4}$. 

\begin{figure}
\centerline{\epsfxsize=2.9in\epsfbox{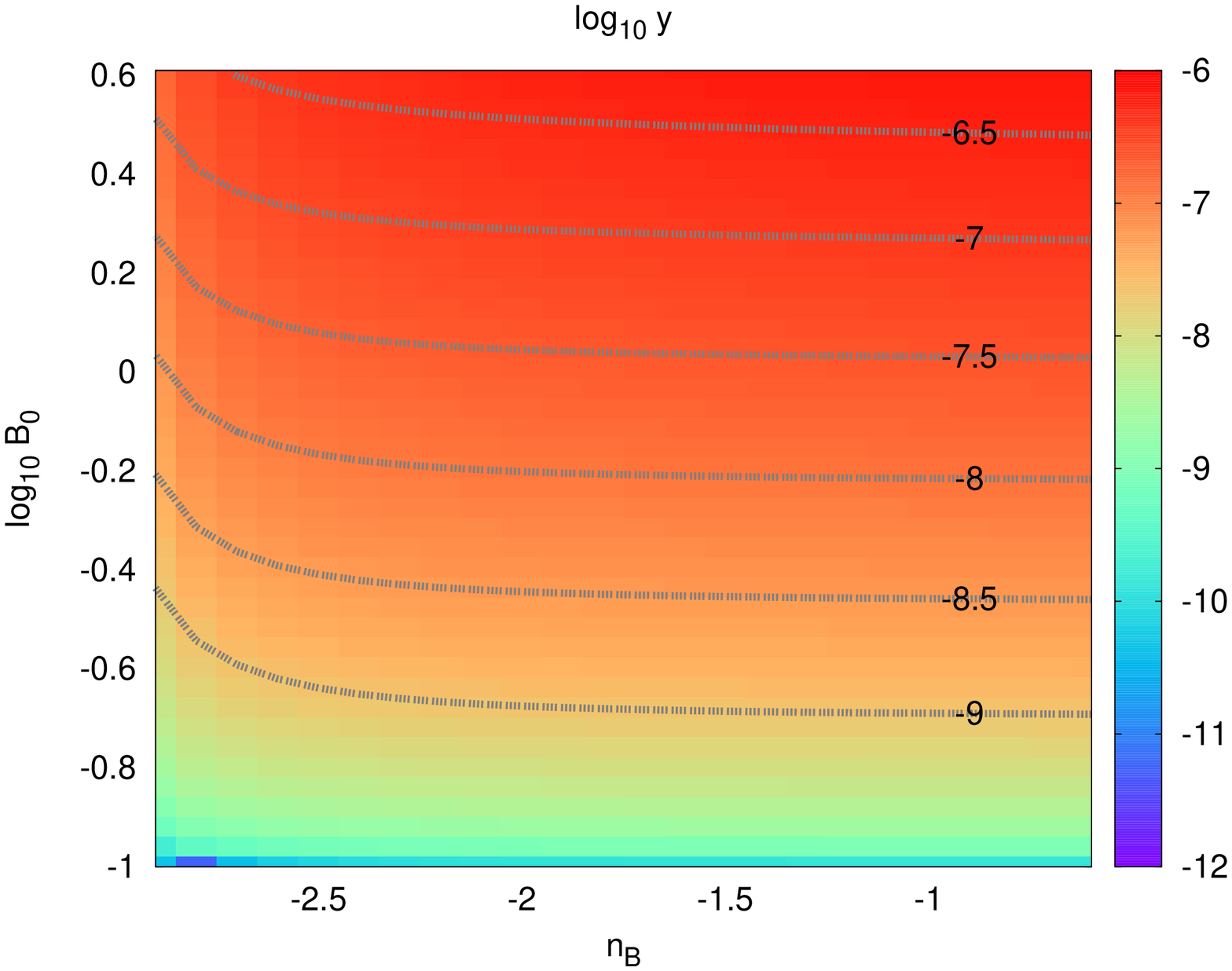}
\hspace{0.9cm}
\epsfxsize=2.9in\epsfbox{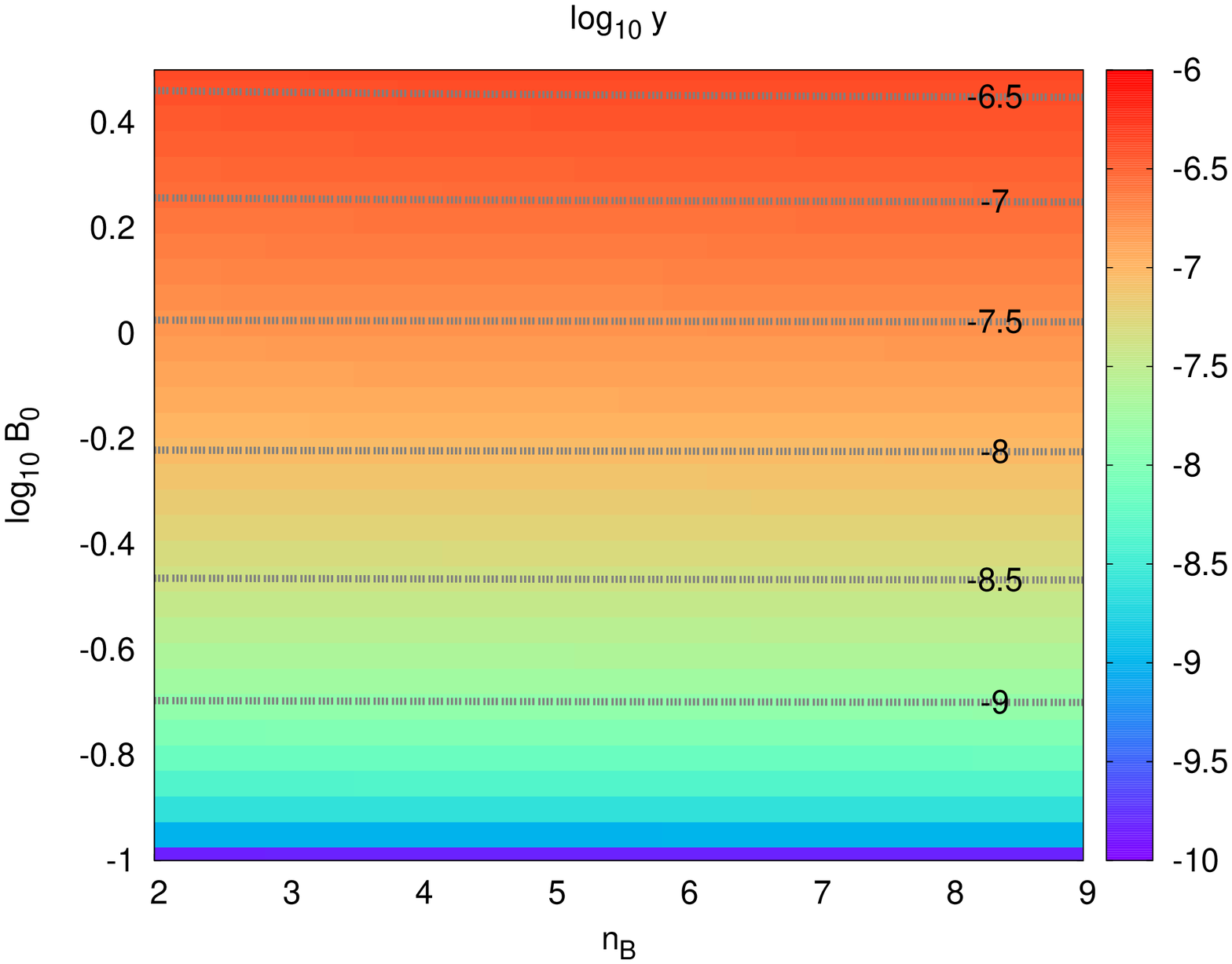}}
\caption{Predicted $y$-type distortion from dissipation of magnetic
 fields in the post-decoupling era due to ambipolar diffusion and
 decaying MHD turbulence as a function of the present-day field value,
 $B_0$, smoothed over $k_{d,dec}$ (eq. (\ref{kdast}) with $\cos\theta=1$),
 and the spectral index, $n_B$. The left panel shows $-2.9\le 
 n_B\le 0$ while the right panel shows $2\le n_B\le 9$.} 
\label{fig13}
\end{figure}

Figure~\ref{fig13} shows the predicted $y$ values for $-2.9\le n_B\le 0$
(left panel) and $2\le n_B\le 9$ (right panel). There are two important
results. First, the predicted values of $y$ are quite sizable, $y\approx
10^{-7}$, for a few nG fields. These values are comparable to the
contributions from virialized halos in $z\lesssim 5$ via the thermal
Sunyaev-Zel'dovich effect, $y_{\rm tSZ}\approx 1.7\times
10^{-6}$ \cite{rksp}, and the reionization of the 
universe at $5\lesssim z\lesssim 10$, $y_{\rm reion}\approx 1.5\times
10^{-7}$. (The 
latter contribution is given 
by $y\approx \tau k_BT_e/(m_ec^2)\approx 1.5\times 10^{-7}$ for
$\tau=0.09$ \cite{wmap9} and $T_e=10^4~{\rm K}=0.862~{\rm eV}$.)

\begin{figure}
\centerline{\epsfxsize=2.9in\epsfbox{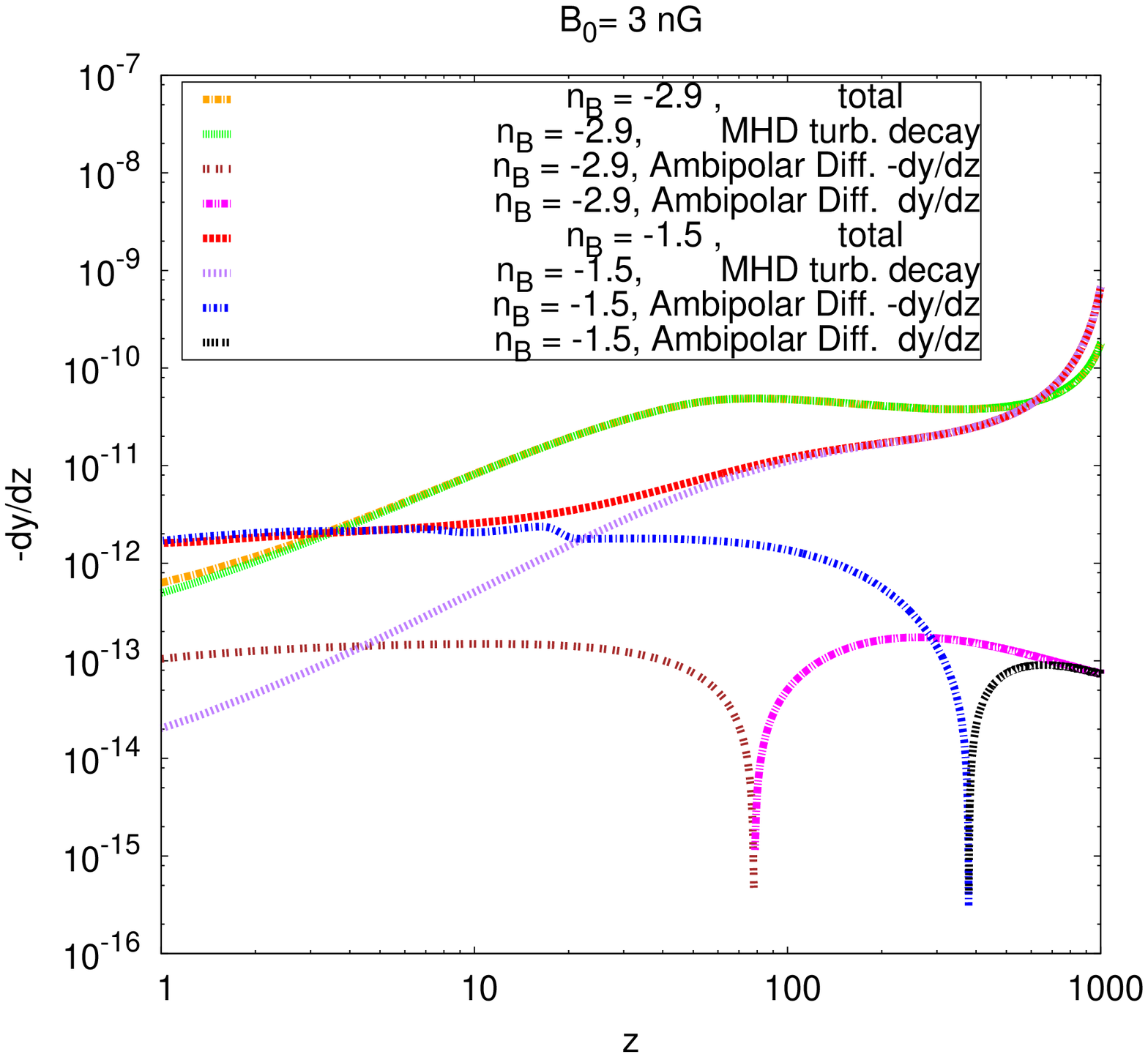}
\hspace{0.9cm}
\epsfxsize=2.9in\epsfbox{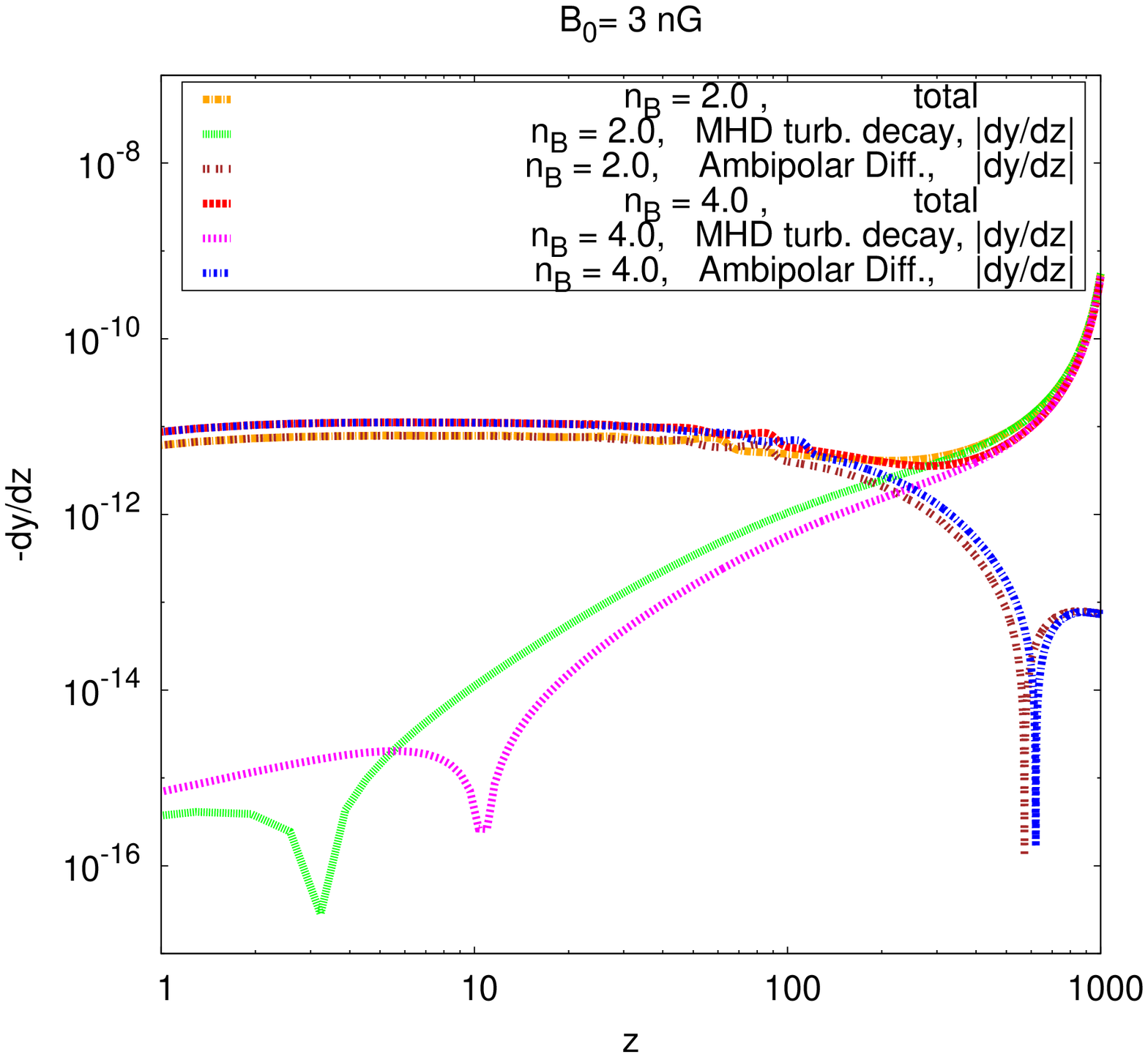}}
\caption{Contribution to the integrand in eq. (\ref{ypost}), $-\frac{dy}{dz}$, from dissipation of magnetic
 fields in the post-decoupling era due to ambipolar diffusion and
 decaying MHD turbulence as a function of the present-day field value,
 $B_0$, smoothed over $k_{d,dec}$ given in eq.~(\ref{kdast}) with $\cos\theta=1$,
 and the spectral index, $n_B$. The left panel shows negative spectral indices
 while the right panel shows positive ones. For all cases, a present-day
 magnetic field of $B_0=3$~nG is assumed. }  
\label{fig13a}
\end{figure}

Second, the predicted values of
$y$ are insensitive to $n_B$ for $n_B\gtrsim -2$ and depend only weakly
on $n_B$ for $-2.9\le n_B\lesssim -2$. 
This is because most of the $y$-type distortion is generated
in early times, $z\gtrsim 100$, when the electron density is high (i.e.,
$n_e\propto x_e(1+z)^3$) and the contribution from  decaying MHD
turbulence completely dominates over that from ambipolar diffusion. For
a given field strength $B_0$, the energy injection from decaying MHD
turbulence depends on $n_B$ only via $m=2(n_B+3)/(n_B+5)$ (see
eq.~\eqref{eq:gammadecay}), which varies slowly with $n_B$ unless $n_B$
is close to $-3$.

Figure~\ref{fig13a} shows the contributions to the integrand of $y$ in
eq. (\ref{ypost}), that is, $-dy/dz$, for  $B_0=3$~nG. 
The $y$-type distortion is completely dominated by decaying MHD
turbulence for $n_B=-2.9$ at all redshifts (see the left panel). As $n_B$
increases, the 
ambipolar diffusion contribution dominates at lower redshifts (see the right
panel). However, dissipation via decaying MHD turbulence continues to be
the dominant contribution at $z\gtrsim 200$ even for $n_B=4$. 
As the dominant contribution to the $y$-type distortion  comes from high
redshifts where the electron density is high, decaying MHD turbulence
always dominates in $y$ for all values of $n_B$.

We find that the COBE/FIRAS bound on $y$ is satisfied for  $B_0<5$~nG
for all values of $n_B$. The expected  PIXIE bound,
$y<10^{-8}$, would exclude $B_0>1.0$ and 0.6~nG for  $n_B=-2.9$ and
$-2.3$, respectively. As the predicted magnitude of $y$ is insensitive
to bluer spectral indices, we find that the expected  PIXIE bound on $y$
would exclude $B_0>0.6$~nG for all values of $n_B\geq 2$. 

\subsection{Optical depth to Thomson scattering}

\begin{figure}
\centerline{\epsfxsize=2.9in\epsfbox{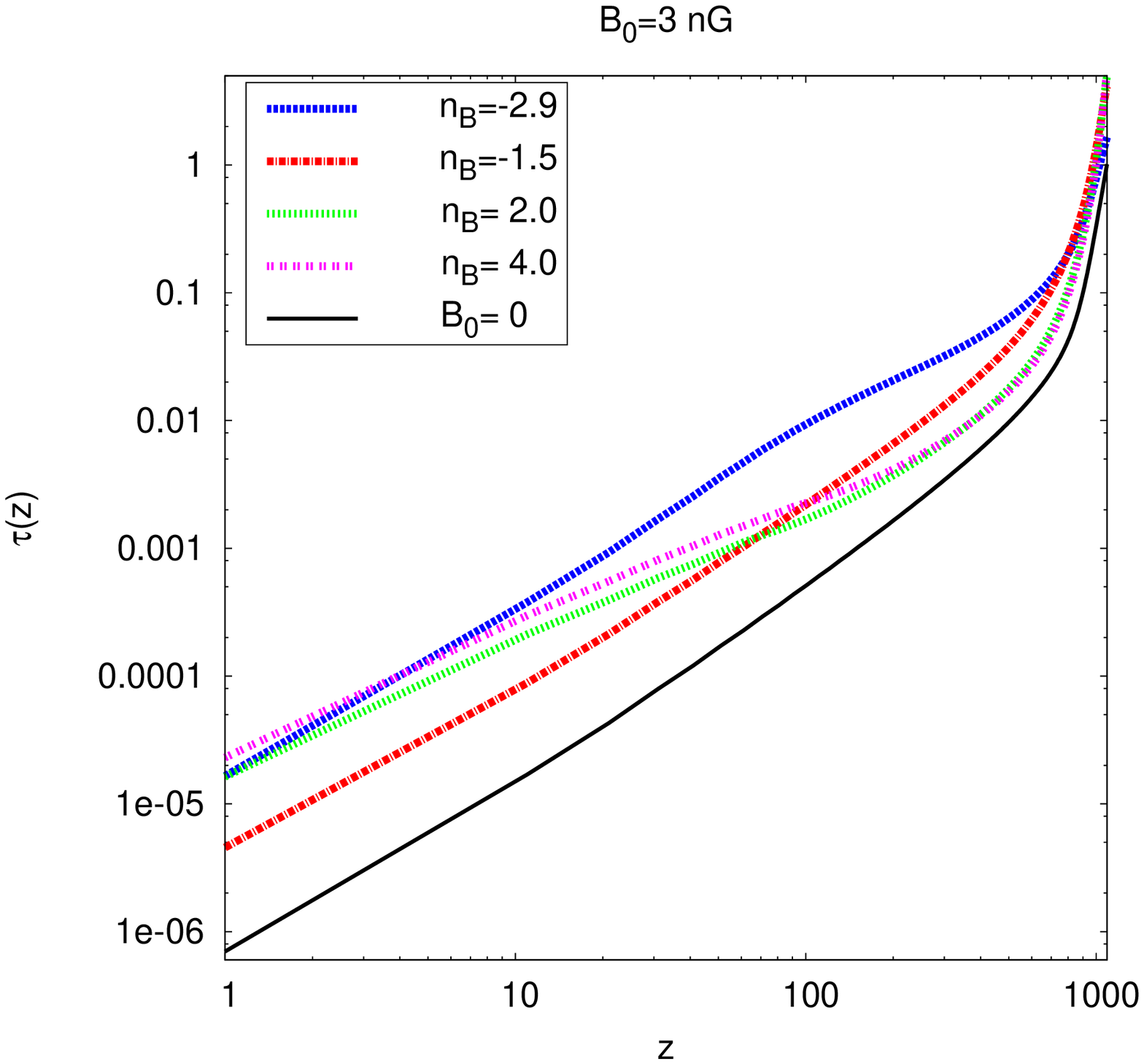}
\hspace{0.9cm}
\epsfxsize=2.9in\epsfbox{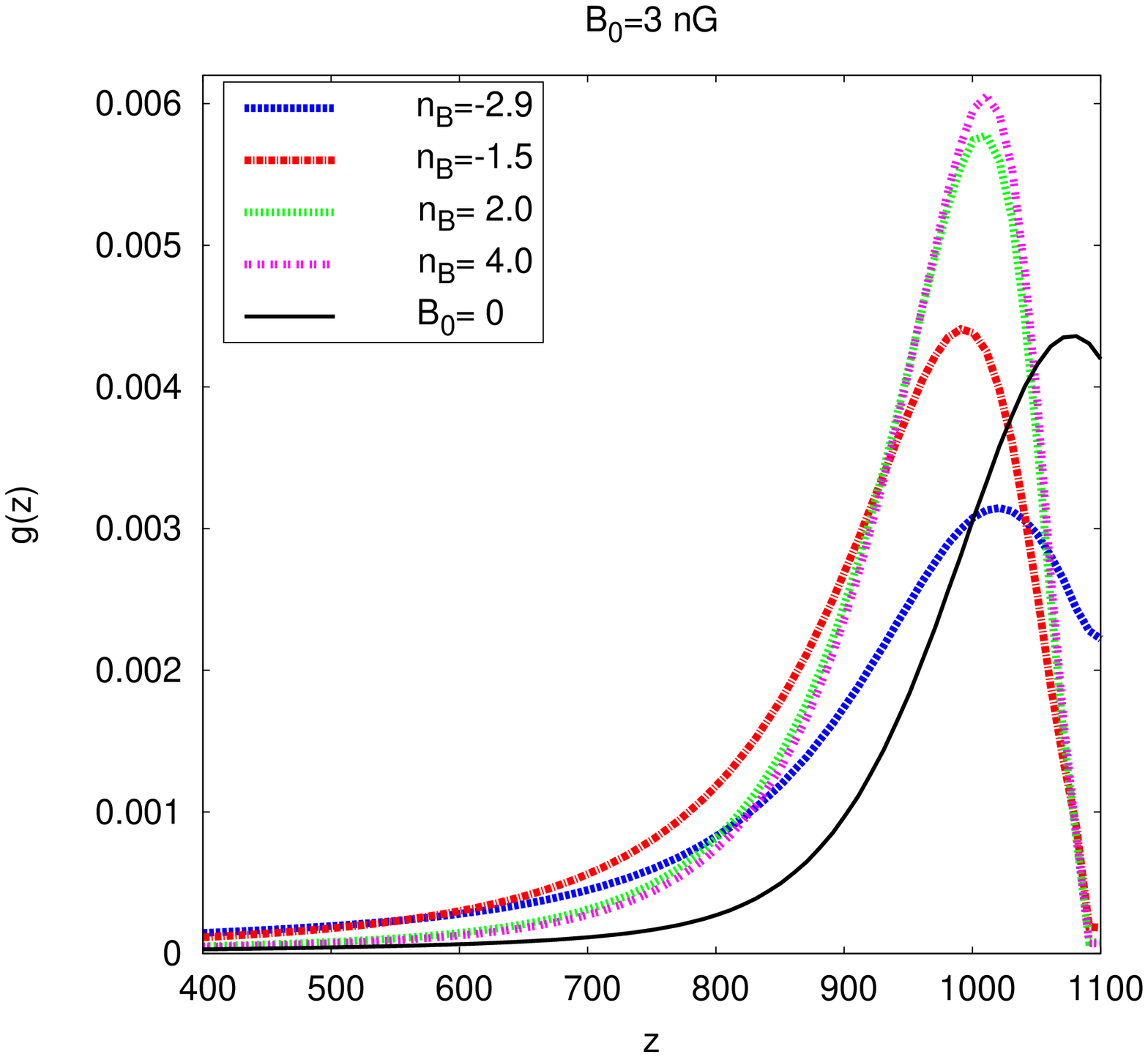}}
\caption{
Optical depth to Thomson scattering from the present epoch to a given
 redshift, $\tau(z)$ (left panel), and the corresponding visibility function,
$g(z)$ (right panel). A present-day magnetic field of $B_0=3$~nG, smoothed over $k_{d,dec}$ (Eq.~(\ref{kdast}) with $\cos\theta=1$),  
 is assumed. Ambipolar diffusion and decaying MHD turbulence are
 included. We explore four different  values of the spectral index
 ($n_B=-2.9$, $-1.5$, $2$, and $4$). The black solid lines show  the case of no
 magnetic field.
}
\label{fig15aa}
\end{figure}

As shown in figure~\ref{fig11}, heating due to dissipation of fields
induces collisional ionization of the IGM. The optical depth to Thomson
scattering resulting from this ionization, integrated from the present
epoch to a given redshift $z$, is given by
\begin{eqnarray}
\tau(z)_{B_0,n_B}=\int_{0}^{z}dz'\frac{\sigma_Tc}{H(z')(1+z')}n_{e,B_0,n_B}(z').
\end{eqnarray}
The left panel of figure \ref{fig15aa} shows $\tau(z)_{B_0,n_B}$ for
$B_0=3~$nG, while the right panel shows the visibility function
defined by $g(z)\equiv \frac{d\tau}{dz}e^{-\tau(z)}$.
Strikingly, even a few nG field yields the optical depth to $z\sim 10^3$
of order unity, which is clearly ruled out by the fact that we can still
see the CMB temperature power spectrum at $l\gtrsim 100$.
Therefore, the optical depth provides an additional constraint on the
field strength \cite{sesu}.

\begin{figure}
\centerline{\epsfxsize=3.5in\epsfbox{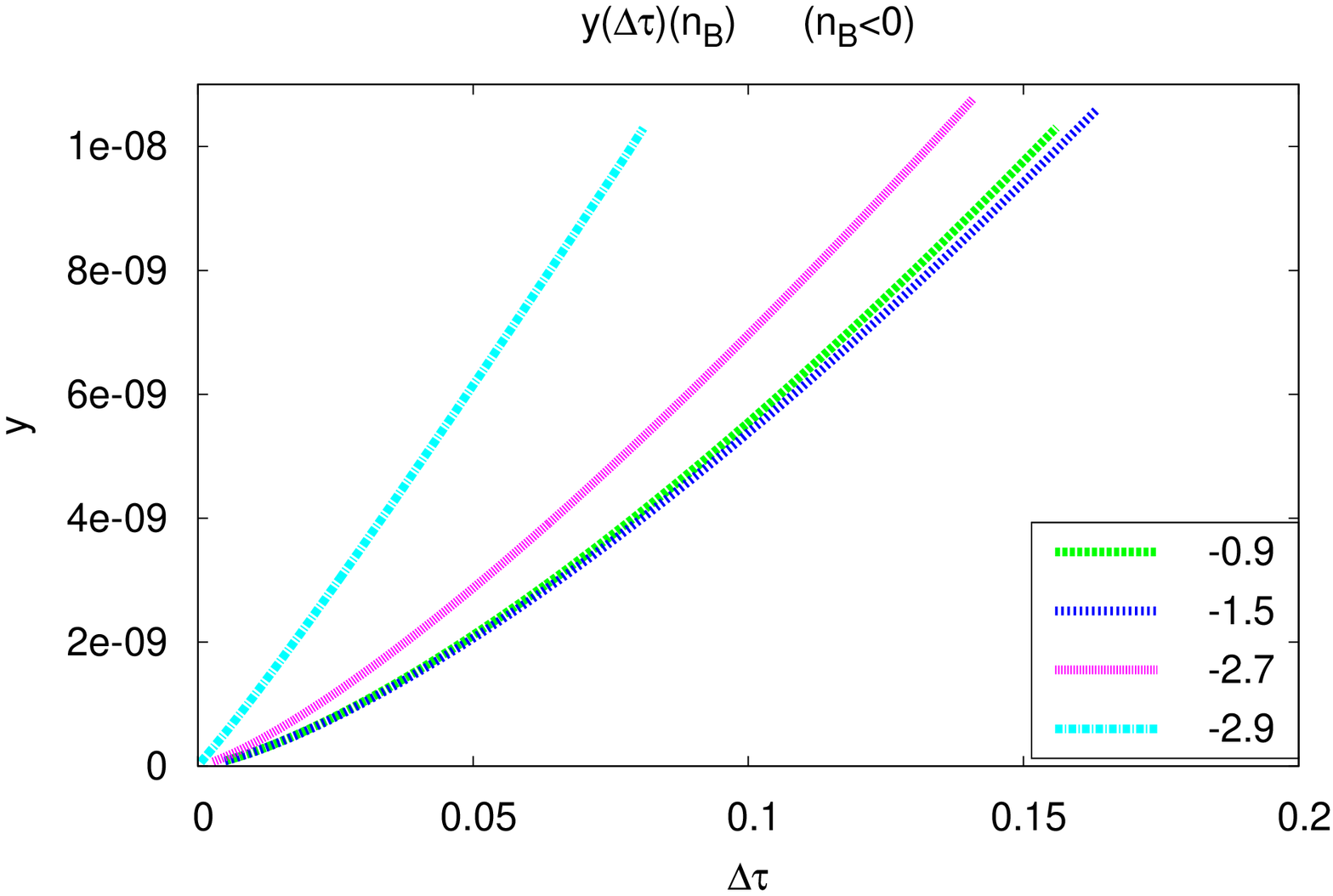}
\hspace{-1cm}
\epsfxsize=3.5in\epsfbox{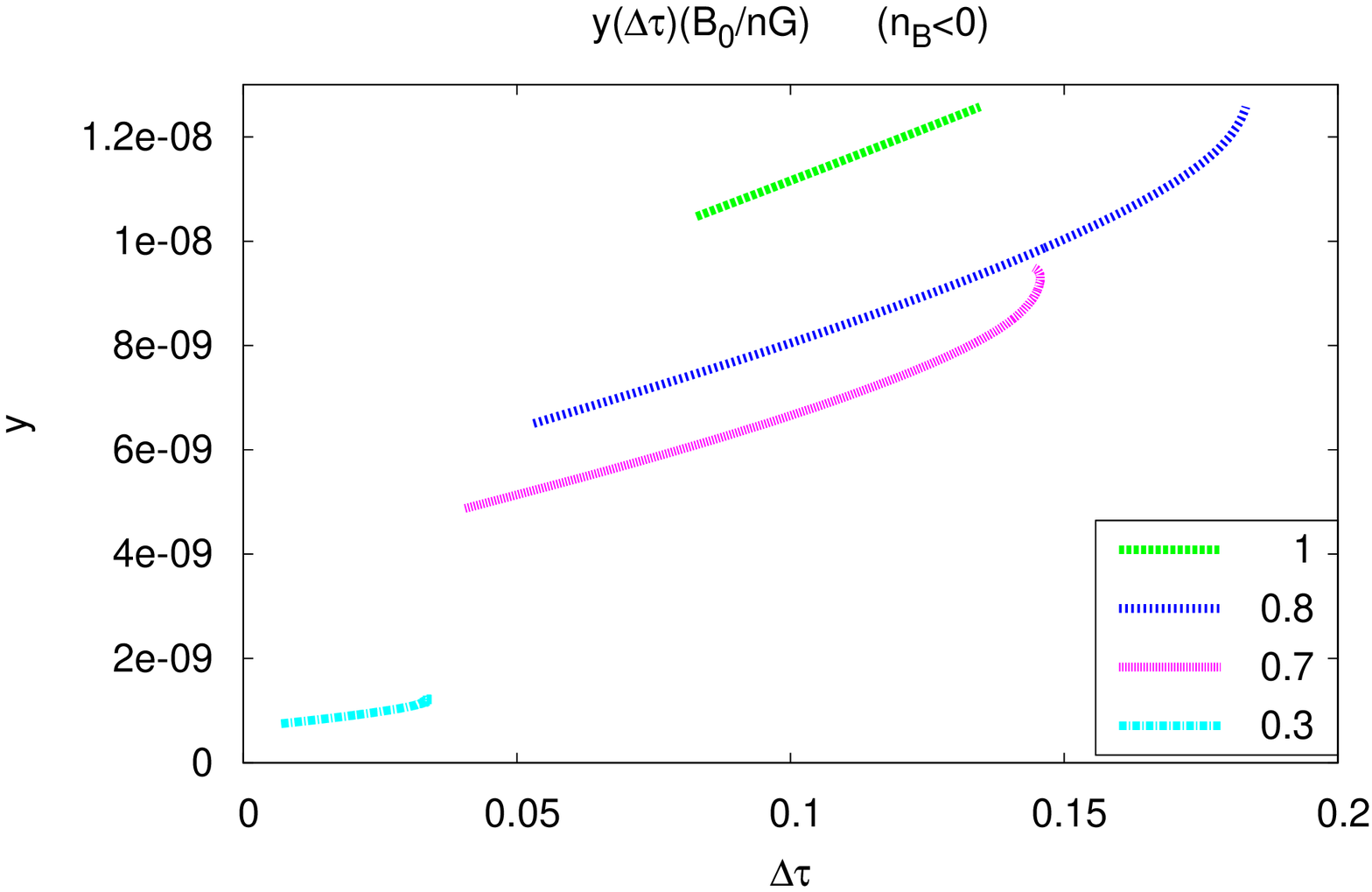}}
\caption{
Correlations between $y$ and $\Delta\tau$ for $n_B<0$. (Left panel) The 
 correlations shown for $n_B=-2.9$, $-2.7$, $-1.5$, and $-0.9$. (The
 field values, $B_0$, vary along each line.)
(Right panel) The correlations shown for $B_0=0.3$, 0.7, 0.8, and
 1~nG. (The spectral indices, $n_B$, vary along each line.)}
\label{fig16aa}
\end{figure}

How much optical depth is allowed by the current CMB data?
The large-scale polarization of CMB at $l\lesssim 10$ constrains the
optical depth up to $z\sim 20$ (see, e.g., ref.~\cite{cmbpol}), while
the CMB temperature power spectrum at $l\gtrsim 100$ constrains the
optical depth all the way up to the decoupling epoch. As almost all of
the optical depth from collisional ionization due to dissipation
of fields is generated at $z\sim 10^3$, the constraint from the
temperature power spectrum is most relevant. However, the
optical depth determined from the temperature power spectrum is
correlated with the amplitude and tilt of primordial fluctuations.

The recently-released {\sl Planck} data are able to break this
correlation between 
parameters, yielding $\tau=0.089\pm 0.032$ (68\%~CL) {\it without using
the polarization data} \cite{Planck_cosmology}. On the other hand, {\sl
WMAP}'s polarization data on large angular scales constrain the optical
depth from reionization of the universe at $z\lesssim 20$ as
$\tau=0.089\pm 0.014$ \cite{wmap9}. Taking the 2-$\sigma$ limits, the
total optical depth may be as high as $\tau_{\rm tot}=0.15$, while the
reionization optical depth may be as low as $\tau_{\rm
reion}=0.061$. Therefore, one may still ``hide'' the optical depth of
$\Delta\tau\sim 0.1$ from a higher-redshift universe.\footnote{However, more
thorough analysis is required if we wish to find precisely how much
optical depth is allowed in a higher-redshift universe, as dissipation
of fields shifts the peak of 
the visibility function to $z\sim 1000$ from $z=1088$ (see the right
panel of figure~\ref{fig15aa}), delaying the
epoch of decoupling. This can modify the CMB temperature power spectrum
significantly.}

We calculate the additional contribution to the optical depth due to the
presence of a large-scale magnetic field between the decoupling and
present epochs as
\begin{eqnarray}
\Delta\tau(B_0,n_B)\equiv\tau_{B_0,n_B}(z_{\rm dec})-\tau_{B_0=0}(z_{\rm dec}).
\end{eqnarray}
For $z_{\rm dec}=1088$, we find convenient numerical fits to our results:
\begin{eqnarray}
\Delta\tau(B_0,n_B)=\left\{
\begin{array}{lr}
0.3(-n_B)^{0.08}\left(\frac{B_0}{\rm nG}\right)^{1.72}
&
\\
-0.03(-n_B)^{0.75}\left(\frac{B_0}{\rm nG}\right)^{1.63}{\rm e}^{1.8\times 10^{-3}(-n_B)^{6.18}},
& n_B<0\\
&\\
&\\
0.27 n_B^{-0.02}\left(\frac{B_0}{\rm nG}\right)^{1.78},
& n_B>0.
\end{array}
\right.
\end{eqnarray}
We also find numerical fits for the $y$-type distortion:
\begin{eqnarray}
y(B_0,n_B)=\left\{
\begin{array}{lr}2.30\times 10^{-8}(-n_B)^{7.8\times 10^{-3}}\left(\frac{B_0}{\rm nG}\right)^{2.42}
&
\\
-1.04\times 10^{-9}(-n_B)^{1.62}\left(\frac{B_0}{\rm nG}\right)^{2.67}{\rm e}^{1.07\times10^{-3}(-n_B)^{6.19}},
& n_B<0\\
&\\
&\\
2.18\times10^{-8} n_B^{0.04}\left(\frac{B_0}{\rm nG}\right)^{2.49},
& n_B>0.
\end{array}
\right.
\end{eqnarray}

\begin{figure}
\centerline{\epsfxsize=3.5in\epsfbox{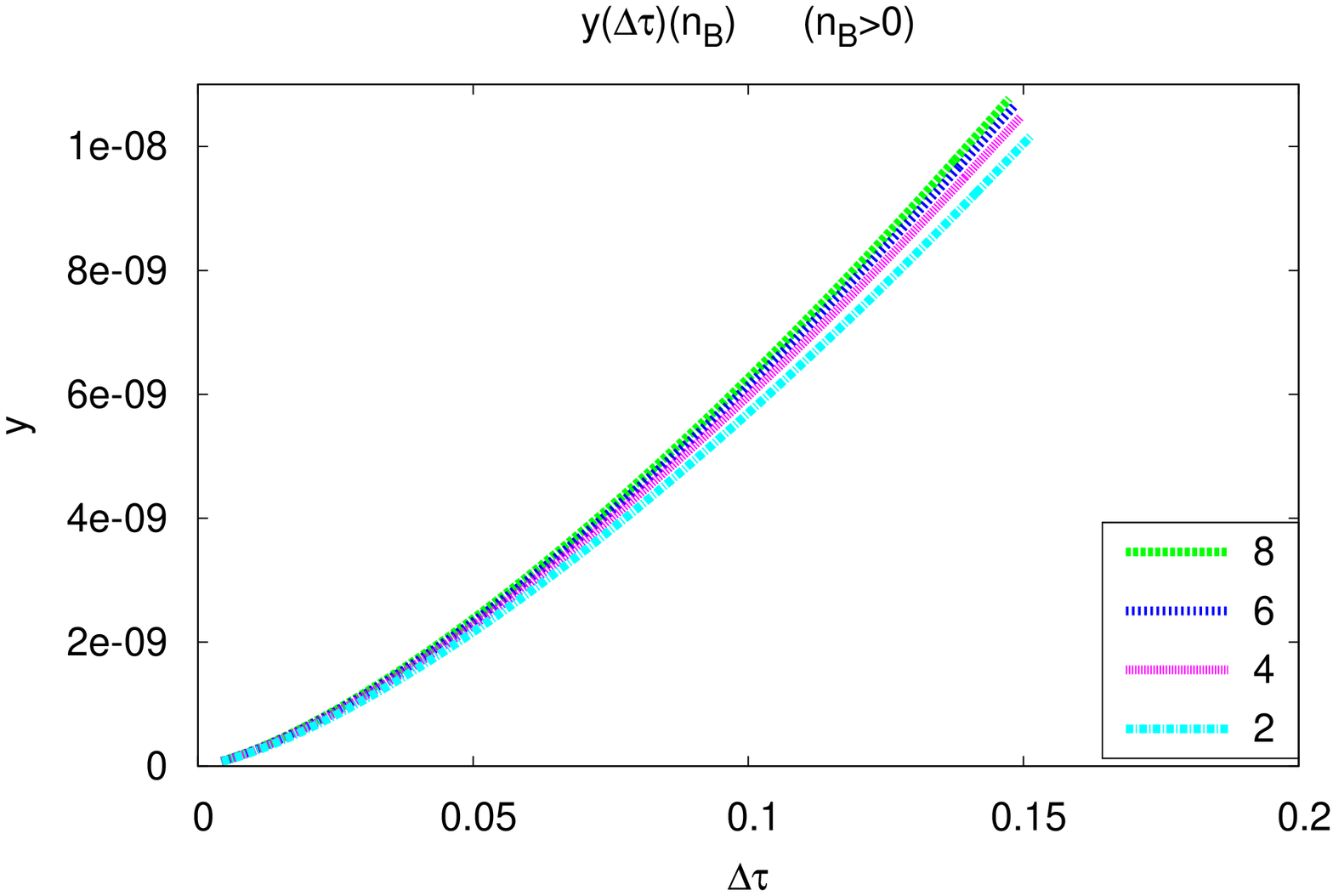}
\hspace{-1cm}
\epsfxsize=3.5in\epsfbox{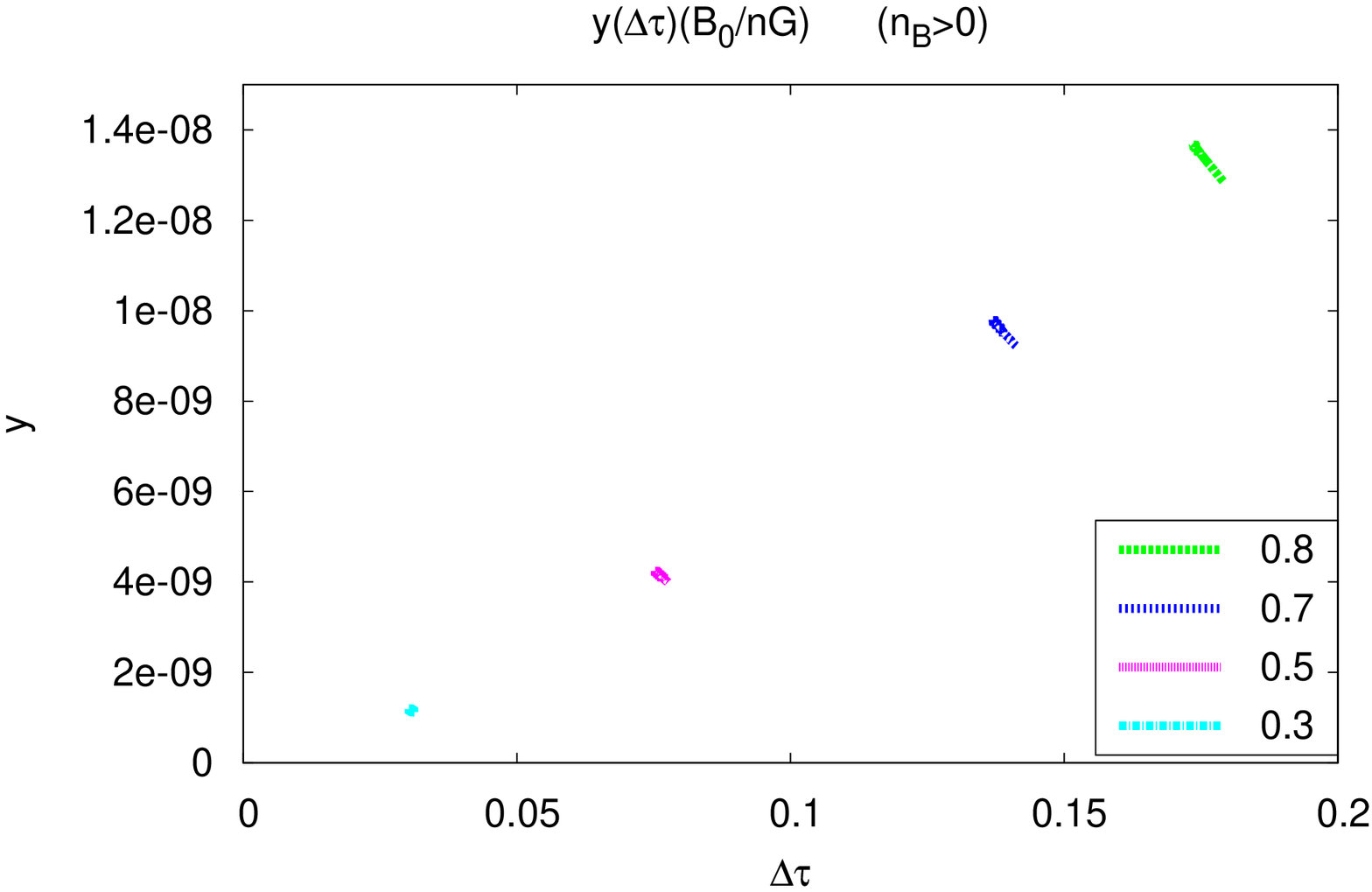}}
\caption{
Correlations between $y$ and $\Delta\tau$ for $n_B\ge 20$. (Left panel) The 
 correlations shown for $n_B=2$, 4, 6, and 8. (The
 field values, $B_0$, vary along each line.)
(Right panel) The correlations shown for $B_0=0.3$, 0.5, 0.7, and
 0.8~nG. (The spectral indices, $n_B$, vary along each line.)}
\label{fig17aa}
\end{figure}

As both $\Delta\tau$ and $y$ are unique functions of $n_B$ and $B_0$,
there is a tight correlation between $\Delta\tau$ and $y$.
We show the correlations for $n_B<0$ and $n_B\ge 2$ in figures
\ref{fig16aa} and \ref{fig17aa}, respectively. We find that
$\Delta\tau\lesssim 0.1$ restricts the predicted $y$-type distortion to
$y\lesssim 10^{-8}$ (with the precise values depending on $n_B$).

\section{Conclusions}
\label{sec:conclusions}

Dissipation of the energy in magnetic fields into the plasma in the
pre-decoupling era as well as that into the IGM in the post-decoupling
era heats photons of the CMB, creating both $y$- and $\mu$-type
distortions of the black-body spectrum of the CMB \cite{jko2,sesu}.

In the pre-decoupling era, fast magnetosonic waves damp at the photon
diffusion scale, $k_\gamma$, whereas slow magnetosonic waves and
Alfv\'en waves damp at significantly a larger wavenumber,
$k_\gamma/v_A\gg k_\gamma$ \cite{jko1,sb}. As a result, if the total
energy is divided 
equally between fast and slow magnetosonic waves and Alfv\'en waves,
dissipation of the energy in the fields is dominated by that of slow
magnetosonic waves and Alfv\'en waves. Dissipation during the
``$\mu$-era,'' $5\times 10^4\lesssim z\lesssim 2\times 10^6$, creates a
$\mu$-type distortion, and that after the $\mu$-era creates a 
$y$-type distortion. 

We find that the $y$- and $\mu$-type distortions from the pre-decoupling
era provide the same limits
on the field strength for a scale-invariant power spectrum of the
fields.  However, the $\mu$-type distortion provides stronger limits for
non-scale-invariant spectra, $n_B\gtrsim -2.9$, as a larger amount of
energy is dissipated at higher redshifts for these spectra.

In the post-decoupling era, the MHD turbulence develops as radiative
viscosity becomes negligible. Non-linear effects then lead to a decay of
the MHD turbulence, leading to dissipation of the magnetic energy. Also,
a separation of charged and neutral particles in the IGM by the Lorentz
force induces a velocity difference between ionized and neutral
components. The magnetic energy is then dissipated via ambipolar diffusion,
as the velocity difference 
is damped by ion-neutral collisions. Both of these effects significantly
alter the thermal and ionization history of the IGM in the
post-decoupling era \cite{sesu}, creating a $y$-type distortion.

We find that decaying MHD turbulence and ambipolar diffusion dominate at
different epochs, with the former dominating at higher redshifts, e.g.,
$z\gtrsim 100$ for $n_B=-1.5$. They have the opposite dependence on
$n_B$: the larger the $n_B$ is, the larger the effect of ambipolar
diffusion becomes, and the smaller the effect of decaying MHD turbulence
becomes. This makes the predicted temperature of the IGM relatively
robust. For the present-day field strength of $B_0=3$~nG smoothed over
$k_{d,dec}$ given by eq.~(\ref{kdast}) with $\cos\theta=1$, we find that
the IGM temperature can rise to $\approx 1000$ to 2600~K at $z\gtrsim 10$ for
all values of $-2.9\le n_B\le 4$. (However, the temperature does not
increase much beyond this for larger field values, as electrons cool via
inverse Compton scattering off the CMB photons.) 
Such a high temperature modifies the
ionization state of the IGM via collisional ionization, yielding the
ionization fraction of order $10^{-3}$ at $z\gtrsim 10$.

The most significant finding in this paper is that a few nG
field smoothed over $k_{d,dec}$ can easily produce $y\approx 10^{-7}$ in
the post-decoupling era. While $y\approx 10^{-7}$ is small, it is well within
sensitivity of the current technology, and a proposed experiment such as
PIXIE would detect $y\approx 10^{-7}$ at the 50-$\sigma$ level
\cite{pixie}. Therefore, such an experiment will be a powerful probe of
the existence of intergalactic (possibly primordial) magnetic fields.

On the other hand, early ionization  of the IGM due to dissipation of fields can
provide a substantial contribution to the optical depth to Thomson
scattering \cite{sesu}. Requiring the additional contribution not to exceed
$\Delta\tau\sim 0.1$ (which seems compatible with the current limits
from the CMB temperature and polarization power spectra), we find that
the predicted $y$-type distortion is restricted to $y\lesssim 10^{-8}$.

A challenge would be to distinguish the $y$-type distortion due to dissipation
of fields from the other known contributions from virialized halos in
$z\lesssim 5$ via the thermal Sunyaev-Zel'dovich effect, $y_{\rm
tSZ}\approx 1.7\times 10^{-6}$ \cite{rksp}, and the reionization of the
universe at $5\lesssim z\lesssim 10$, $y_{\rm reion}\approx 1.5\times
10^{-7}$. The thermal Sunyaev-Zel'dovich effect is correlated with the
large-scale structure of the universe traced by galaxies and galaxy
clusters, whereas the reionization signal is correlated with 21-cm lines
from neutral hydrogen during the reionization epoch. These correlations
may be used to distinguish between the sources of the post-decoupling
$y$-type distortion.

\section{Acknowledgements}
We would like to thank Jens Chluba and Rishi Khatri for useful discussion.
KEK  would like to thank the Max-Planck-Institute for Astrophysics for
hospitality where  this work was initiated. KEK acknowledges financial
support by Spanish Science Ministry grants FIS2012-30926 and
CSD2007-00042. We acknowledge the use of the Legacy Archive for
Microwave Background Data Analysis (LAMBDA). Support for LAMBDA is
provided by the NASA Office of Space Science.

\appendix
\section{Photon diffusion scale}
\label{sec:appendix}

Including polarization, the photon diffusion scale is given by \cite{kaiser}
\begin{eqnarray}
k_{\gamma}^{-2}(z)=\int_z^{\infty}\frac{dz}{6H(z)(1+R)\dot{\tau}}\left(\frac{16}{15}+\frac{R^2}{1+R}\right).
\label{exact}
\end{eqnarray}
The baryon-to-photon density ratio, $R$, is given by
 $R=\frac{3}{4}\frac{\Omega_{b,0}}{\Omega_{\gamma,0}}(1+z)^{-1}$. For
 the best-fit parameters of the {\sl WMAP} 9-year data only
 ($\Omega_{b,0}h^2=0.02264$ and $\Omega_{\gamma,0}h^2=2.471\times
 10^{-5}\left(\frac{T}{2.725 {\rm K}}\right)^4$ \cite{pdg}), the
 numerical pre-factor evaluates to 687.171. Since $z>1000$, it is a good
 approximation to neglect $R$.
For the expansion rate, we use $H(z)=H_0\Omega_{r,0}^{\frac{1}{2}}\left(1+z\right)^2\left(1+\frac{1+z_{eq}}{1+z}\right)^{\frac{1}{2}}$ where 
$\Omega_{r,0}=1.69\, \Omega_{\gamma,0}$ is the present-day total density
 of relativistic species including the 
standard value for the  effective number of light neutrinos,
 $N_\nu=3.04$. The epoch of radiation-matter equality is given by
 $\Omega_{r,0}=\Omega_{m,0}/(1+z_{eq})$. 
Thus
\begin{eqnarray}
k_{\gamma}^{-2}\simeq\frac{8}{45}H_0^{-1}\Omega_{r,0}^{-\frac{1}{2}}\int_{z}^{\infty}\frac{dz}{z^2\left(1+\frac{z_{eq}}{z}\right)^{\frac{1}{2}}}\frac{1}{\dot{\tau}}.
\label{app0}
\end{eqnarray}
Following \cite{HuSu} the differential optical depth can be approximated by
\begin{eqnarray}
\dot{\tau}(z)=\frac{c_2}{1000}\Omega_b^{c_1}\left(\frac{z}{1000}\right)^{c_2-1}\frac{\dot{a}}{a}\left(1+z\right),
\label{app1}
\end{eqnarray}
where a dot indicates the derivatives w.r.t. conformal time, $c_1=0.43$,
and $c_2=16+1.8\ln\Omega_b$. 
We compute  the ionization fraction using $x_e(z)={\rm
min}(\dot{\tau}(n_e\sigma_T\frac{a}{a_0})^{-1},1)$, where
\begin{eqnarray}
\left(n_e(z)\sigma_T\frac{a}{a_0}\right)^{-1}=4.34\times 10^4\left(1-\frac{Y_p}{2}\right)^{-1}\left(\Omega_bh^2\right)^{-1}\left(\frac{T}{2.725 {\rm K}}\right)^{-3}\left(1+z\right)^{-2}\;\; {\rm Mpc}.
\end{eqnarray}
For the best-fit parameters of the {\sl WMAP} 9-year data only, $x_e$
calculated using the expression (\ref{app1}) is larger than one for
$z>z_{*}\simeq 1486.57$. Thus, the differential optical depth is given
by eq. (\ref{app1}) for $z_{\rm dec}<z<z_*$ and by
$\dot{\tau}=n_e\sigma_T\frac{a}{a_0}$ for $z\geq z_*$. 

\begin{figure}
\centerline{\epsfxsize=2.9in\epsfbox{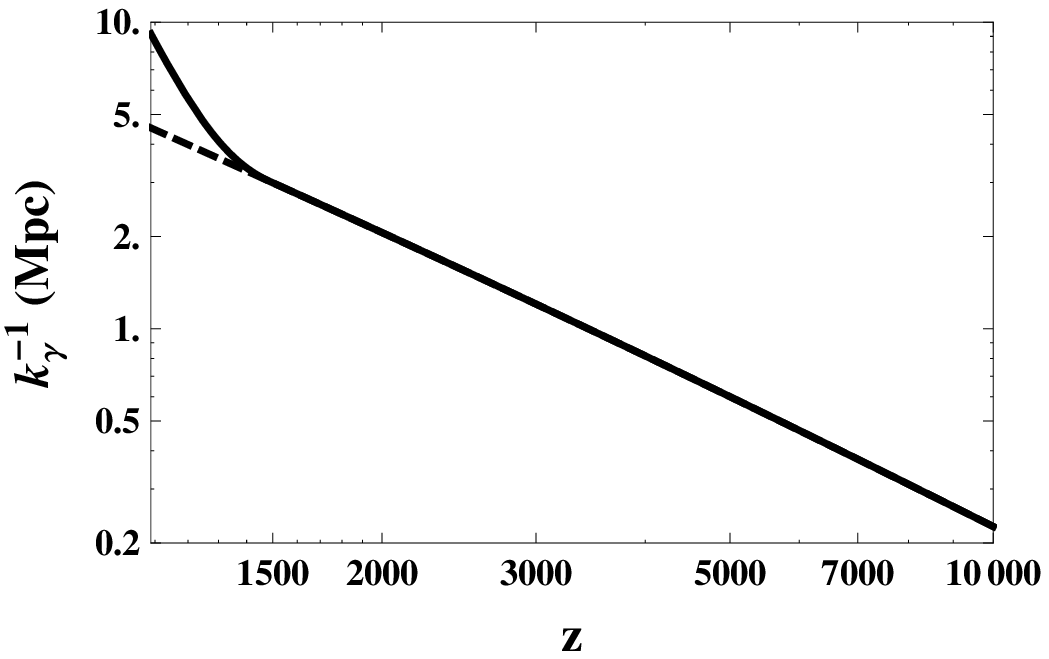}
\hspace{0.9cm}
\epsfxsize=2.9in\epsfbox{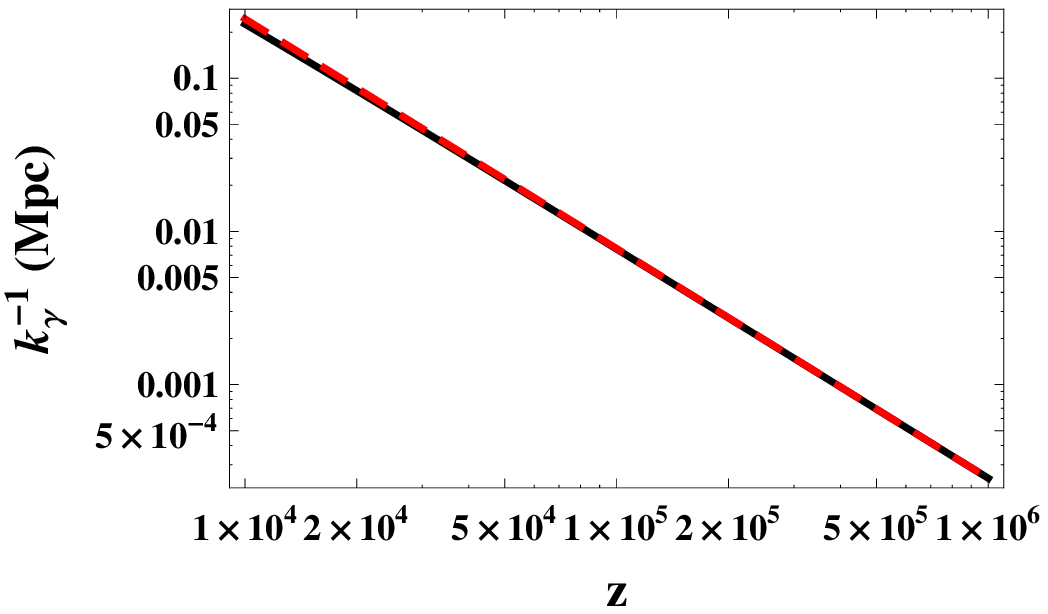}}
\caption{Photon diffusion scale as a function of $z$. (Left panel) For
 $z<10^4$. The
 solid line shows the results using eqs.~(\ref{numk.z.gt.z*}) and
 (\ref{numk.z.lt.z*}). The dashed line shows the photon diffusion scale
 for instantaneous recombination.  (Right panel) For $z>10^4$.
 The solid (black) line shows the results using
 eq.~(\ref{numk.z.gt.z*}), while the dashed (red) line shows the
 approximation in the radiation-dominated era, eq.~(\ref{kgaRad}).}
\label{fig15}
\end{figure}

Deep inside the radiation-dominated era, the photon diffusion scale approaches
\begin{eqnarray}
k_{\gamma}^{-2}\to A_{\gamma}^2z^{-3},
\label{kgaRad}
\end{eqnarray}
where $A_{\gamma}^2=5.9807\times 10^{10}\;{\rm Mpc}^2$  for the best-fit
parameters of the {\sl WMAP} 9-year data only.
This is compared with the exact numerical result in the right panel of
figure \ref{fig15}.

For completeness we give the expressions for $k_\gamma^{-2}$ in  $z\geq z_*$,
\begin{eqnarray}
k_{\gamma, \; z\geq z_*}^{-2}(z)&=&2.16567\times 10^7\left(\Omega_{r,0}h^2\right)^{-\frac{1}{2}}\left(1-\frac{Y_p}{2}\right)^{-1}\left(\Omega_b h^2\right)^{-1}
\nonumber\\
&\times&
\int_z^{\infty} dz (1+z)^{-\frac{7}{2}}(2+z+z_{eq})^{-\frac{1}{2}}(1+R)^{-1}\left(\frac{16}{15}+\frac{R^2}{1+R}\right) \;\; {\rm Mpc}^2
\label{numk.z.gt.z*}
\end{eqnarray}
and in $z_{\rm dec}<z<z_*$,
\begin{eqnarray}
k_{\gamma, \;z<z_*}^{-2}(z)&=&1.49402\times 10^6(\Omega_{r,0}h^2)^{-1}\Omega_b^{-c_1}c_2^{-1}10^{3c_2}
\nonumber\\
&\times&\int_{z}^{z_*}dz(1+z)^{-3}(2+z+z_{eq})^{-1}z^{1-c_2}(1+R)^{-1}\left(\frac{16}{15}+\frac{R^2}{1+R}\right) \;\; {\rm Mpc}^2
\nonumber\\
&+&k_{\gamma, \;z\geq z_*}^{-2}(z_*).
\label{numk.z.lt.z*}
\end{eqnarray}
The evolution of $k_{\gamma}^{-1}$ is shown in the left panel of figure
\ref{fig15}, which clearly shows the significant increase in the photon
mean free path close to decoupling. 

For the parameters of the {\sl WMAP} 9-year data only, the maximal
damping wave number at decoupling computed from eq.~(\ref{numk.z.lt.z*}) is
\begin{eqnarray}
k_{d,dec}=286.91\left(\frac{B_0}{\rm nG}\right)^{-1}\;\; {\rm Mpc}^{-1}.
\end{eqnarray}
\bibliography{references}
\end{document}